\documentclass[%
 reprint,
 showpacs,
 amsmath,amssymb,
 aps,
 pra,
]{revtex4-1}

\usepackage[ascii]{inputenc}
\usepackage[T1]{fontenc}
\usepackage[english]{babel}
\usepackage{graphicx}
\usepackage{microtype}
\usepackage{bm}
\usepackage[pdftex,colorlinks]{hyperref}
\hypersetup{%
  linkcolor=blue,%
  citecolor=blue,%
  urlcolor=blue,%
  pdftitle={PT-Symmetric gain and loss in a rotating frame},%
  pdfauthor={Daniel Haag, Dennis Dast,
    Holger Cartarius, G\"unter Wunner}%
}

\newcommand{\PT}{\mathcal{PT}}

\newcommand{\mrm}{\mathrm}
\newcommand{\rmi}{\mathrm{i}}
\newcommand{\rme}{\mathrm{e}}
\newcommand{\rmd}{\mathrm{d}}

\begin{document}

\title{$\PT$-symmetric gain and loss in a rotating Bose-Einstein condensate}

\author{Daniel Haag}
\email[]{daniel.haag@itp1.uni-stuttgart.de}

\author{Dennis Dast}

\author{Holger Cartarius}

\author{G\"unter Wunner}

\affiliation{Institut f\"ur Theoretische Physik 1,
  Universit\"at Stuttgart, 70550 Stuttgart, Germany}

\date{\today}

\begin{abstract}
  $\PT$-symmetric quantum mechanics allows finding stationary states in
  mean-field systems with balanced gain and loss of particles.
  In this work we apply this method to rotating Bose-Einstein condensates with
  contact interaction which are known to support ground states with vortices.
  Due to the particle exchange with the environment transport phenomena
  through ultracold gases with vortices can be studied.
  We find that even strongly interacting rotating systems support stable
  $\PT$-symmetric ground states, sustaining a current parallel and
  perpendicular to the vortex cores.
  The vortices move through the non-uniform particle density and leave or
  enter the condensate through its borders creating the required net current.
\end{abstract}

\maketitle

\section{Introduction}
\label{sec:Introduction}
One of the most desired effects studied in Bose-Einstein condensates is the
formation of vortices that arise when the condensates are brought into
rotation~\cite{Fetter09a}.
Vortices were first studied in different low-temperature quantum systems
such as superfluid helium~\cite{Feynman55a} and
superconductors~\cite{Abrikosov57a}.
In 1961, the Gross-Pitaevskii equation was formulated to describe vortices in a
Bose-Einstein condensate~\cite{Pitaevskii61a, Gross61a}.
Shortly after the first realizations of these ultracold
condensates~\cite{Davis95a, Anderson95a}, Butts and Rokhsar~\cite{Butts99a}
used a variational approach consisting of a linear combination of the
low-energy angular-momentum eigenstates of the harmonic oscillator to show that
such vortices form themselves if the Bose-Einstein condensate is rotated.
Only months later the calculations where experimentally confirmed by Madison
et al.~\cite{Madison00a}.
At this time, numerical methods such as the finite-element method made the
exact numerical study of these systems possible, thus providing a much higher
precision~\cite{Bao07a}.

The study of transport phenomena is a common interest in Bose-Einstein
condensates, e.g., through optical lattices~\cite{Choi99a} or even random
potentials~\cite{Clement05a}, as well as in superfluids~\cite{Putterman74a,
Pethick77a}, and, naturally, in superconductors~\cite{Campbell72a}.
If a Bose-Einstein condensate is discussed in the mean-field approximation of
the Gross-Pitaevskii equation, the necessary particle gain and loss can be
described by imaginary potentials~\cite{Kagan98a}, rendering the Hamiltonian
non-Hermitian~\cite{Moiseyev11a}.
Up to now, such Hamiltonians have been studied extensively~\cite{Kagan98a,
Schlagheck06a, Rapedius09a, Rapedius10a, Abdullaev10a, Bludov10a, Witthaut11a}
and the particle in- and out-coupling were compared to many-particle
calculations justifying their use in mean-field theory~\cite{Rapedius13a,
Dast14a}.
The particle loss can be induced by a focused electron beam~\cite{Gericke08a}
while particles are added by letting them fall into the condensate from a
second condensate~\cite{Robins08a}.

In 1998, Bender and Boettcher~\cite{Bender98a} discovered that
non-Hermitian Hamiltonians can support stationary solutions if they are
$\PT$ symmetric.
This finding not only opened the possibility of postulating new theoretical
concepts to replace the long accepted requirement of Hermitian
Hamiltonians~\cite{Bender02a,Mostafazadeh08a,Mostafazadeh10a} but started many
other theoretical and experimental studies in optical~\cite{El-Ganainy07a,
  Klaiman08a, Musslimani08a, Makris08a, Makris10a, Guo09a, Ruter10a, Peng14b,
Regensburger12a} and electronic systems~\cite{Schindler11a, Bender13a}.
In these systems, the non-Hermitian Hamiltonian does not describe the full
quantum mechanical system but a macroscopic quantity such as the electric field
in a wave guide system or the electric current.
Inspired by the proposal of Klaiman et al.~\cite{Klaiman08a}, Bose-Einstein
condensates were studied as an additional realization using a double-well
system, where particles are injected into one well and removed from the
other~\cite{Cartarius12a, Cartarius12b, Dast13a, Dast13b,
Haag14b}.

These studies show that $\PT$-symmetric condensates provide an excellent
framework for the theoretical study of particle transport through vortex
systems.
There exists a vast amount of numerical and exact analytical calculations,
describing stationary vortex states in $\PT$-symmetric
systems~\cite{Achilleos12a, Konotop16a, Schwarz17a}.
However, these vortex states are not stable ground states but instead highly
excited states.
Due to the $\PT$-symmetric potential, the vortex structure of such states is
typically lost for strong particle in- and out-coupling.

The formation of vortices in the ground state can be achieved by rotating a
non-isotropic trap~\cite{Landau79a}, stirring the condensate~\cite{Madison00a},
or inducing a synthetic magnetic field~\cite{Lin09a, Zhao15a}.
The different methods all lead to similar equations of motion~\cite{Landau79a,
Madison00a, Zhao15a}, one of which is the Gross-Pitaevskii equation in a
rotating frame.
In natural units and for a rotation axis and angular frequency $(0, 0,
\Omega)$ it reads
\begin{align}
  \label{eq:GPEtimeRotatingFrame}
  \rmi\frac{\partial}{\partial t} &\psi \left( \bm{r}, t \right) =  \\
  &\left(- \Delta + V_{\mrm{rot}}\left( \bm{r} \right) + 8 \pi N a \left|
  \psi\left( \bm{r}, t \right) \right|^2 - \Omega \hat{L}_z \right) \psi
  \left( \bm{r}, t \right),\notag
\end{align}
where $V_{\mrm{rot}}$ describes the rotating potential in the rotating frame.

A solution $\psi \left( \bm{r}, t \right) = \psi \left( \bm{r}
\right) \rme^{-\rmi \mu t}$ is stationary with respect to the rotating frame
and fulfills the stationary rotating Gross-Pitaevskii equation,
\begin{align}
  \label{eq:GPERotatingFrame}
  \mu & \psi \left( \bm{r} \right) =  \\
   &\left(- \Delta + V_{\mrm{rot}}\left(
  \bm{r} \right) + 8 \pi N a \left| \psi\left( \bm{r} \right) \right|^2 -
  \Omega \hat{L}_z \right) \psi \left( \bm{r} \right).\notag
\end{align}
Note that for $t = 0$ the wave functions in the rotating and laboratory
frame are the same.
It can be directly seen that the mean-field energy in the rotating frame
differs from the non-rotating form, $E_\mrm{MF}$, and reads
\begin{equation}
  E_\mrm{MF, rot} = E_\mrm{MF} - \Omega \langle \hat{L}_z \rangle.
  \label{eq:EMFRot}
\end{equation}
If the potential lacks complete rotational symmetry, a non-rotating state can
no longer be stationary.
The ground state of the system is then determined by the modified mean-field
energy $E_\mrm{MF, rot}$~\cite{Landau79a}, allowing the study of vortex
filaments in superfluids and other coherent quantum material.

Before discussing our numerical results, we have to emphasize that since the
potential is time independent in the rotating frame it rotates with the same
frequency as the condensate.
In particular in the case of potentials that are not isotropic in the rotating
plane, this must be considered.
We use an ansatz of two- and three-dimensional B-splines for the algorithm
described in~\cite{Haag15a}.
To achieve more accurate results, the density of elements in the coordinate
space are chosen recursively such that the error of the kinetic part is
minimized.
In Sec.~\ref{sec:2D} a two-dimensional study of a rotating $\PT$-symmetric
potential is studied.
To study particle transport along the vortices, a three-dimensional study is
performed in Sec.~\ref{sec:3D}, followed by a short conclusions in
Sec.~\ref{sec:Conclusion}.

\section{Two-dimensional system}
\label{sec:2D}
Due to the enormous numerical advantage it is reasonable to start with
two-dimensional calculations, i.e., the lowest dimensional system that can
provide states with vortices.
To illustrate the reduction from three to two dimensions consider a harmonic
oscillator with the radial trapping frequency $\omega_r = 1$ and the axial
trapping frequency $\omega_z$,
\begin{equation}
  V(r, z) = \frac{1}{4} r^2 + \frac{1}{4} \omega_z^2 z^2.
  \label{eq:harmonicOscillatorRZPotReal}
\end{equation}
For a linear system a separation into the radial and the axial parts is
possible.
Due to strong axial trapping frequencies $\omega_z \gg 1$ excitations in this
direction are suppressed and the wave function assumes the shape of the
harmonic oscillator ground state in that direction.

The interaction energy in the remaining two-di\-men\-sio\-nal equation, $\int
\mrm{d}x \mrm{d}y 8\pi N a |\psi(x,y)|^4$, equals the interaction energy in the
original three-dimensional system if one assumes $\omega_z \approx 4 \pi$ for
the trapping in $z$-direction. 
In this case, the rotating Gross-Pitaevskii equation in two dimensions assumes
the form
\begin{align}
  \label{eq:toSolveRot2D}
  \mu& \psi(x,y) =\\
  &\left(-\Delta + \frac{1}{4} (x^2 + y^2) + 8 \pi N a
  |\psi(x,y)|^2 - \Omega \hat{L}_z \right) \psi(x,y),\notag
\end{align}
where the interaction strength $8\pi Na$ appears unmodified.
For stronger repulsive contact interactions this approximation is not exact
and an even tighter trapping in $z$-direction must be employed.
However, in this section, an adequately strong trapping frequency is assumed
and the interaction strengths are fixed.

To describe the in- and out-coupling of particles the $\PT$-symmetric
imaginary potential
\begin{equation}
  V_I(y) = - \rmi \gamma\, \text{sign}(y)
  \label{eq:harmonicOscillatorImag2D}
\end{equation}
is used.
For positive values of $\gamma$ the potential describes a gain of particles for
$y < 0$ and a loss of particles for $y > 0$.
The potential is constant
in both regions.
This ensures that the same amount of particles is coupled in and out of the
system for every possible $\PT$-symmetric wave function.
Thus, a comparison between different parameter sets and different numbers
of vortices is directly possible.
It must be emphasized that the potential is used inside the rotating frame and
therefore is itself rotating around the point $x=y=0$, i.e., the gain and loss
contributions are rotating alongside the wave function.

\begin{figure} 
  \centering
  \includegraphics{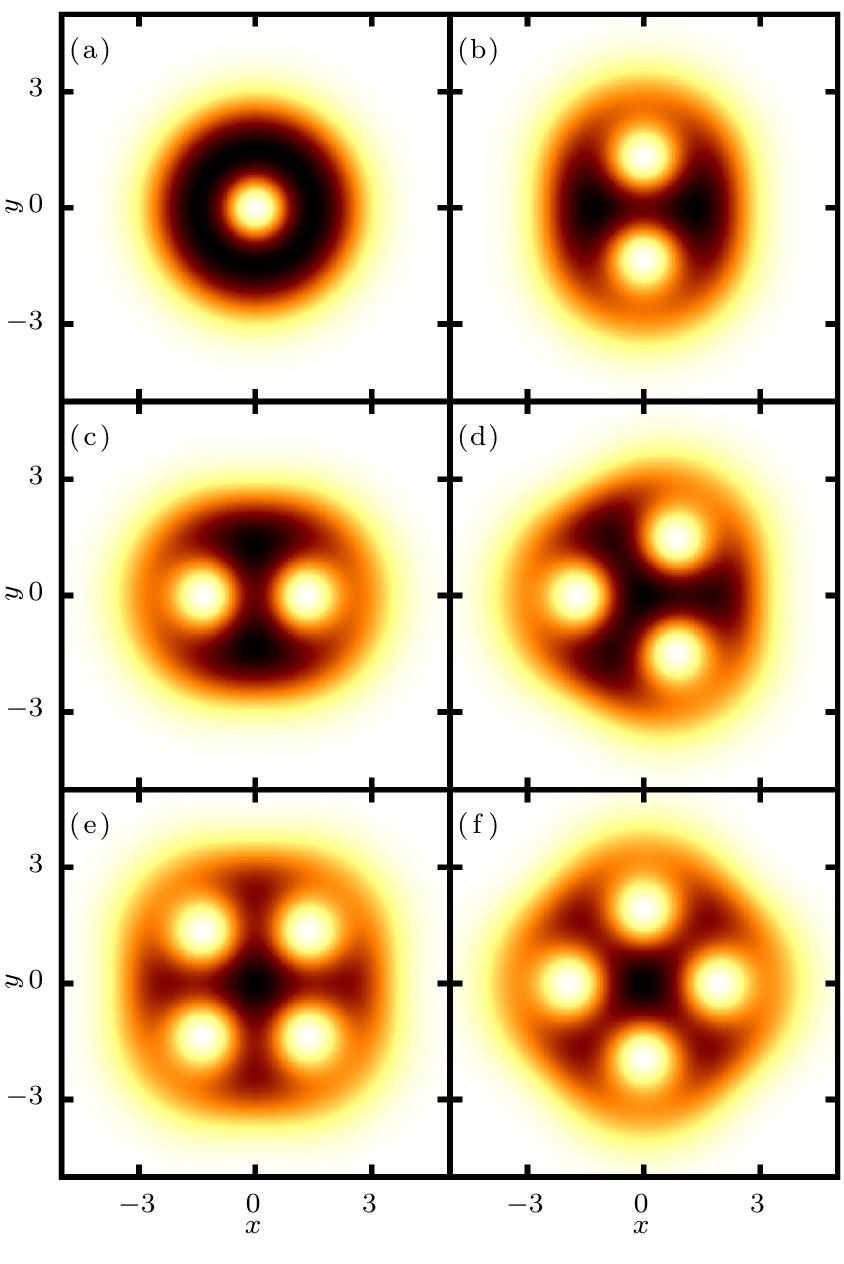}%
  \caption{
    Ground states of the rotating isotropic harmonic oscillator without gain
    and loss ($\gamma = 0$) for $\Omega = 0.85$ (a), $\Omega = 0.88$ (b),(c),
    $\Omega = 0.91$ (d), $\Omega = 0.94$ (e),(f) in all possible $y$-symmetric
    configurations.
    Note that the states with two, three and four vortices exhibit a two-,
    three- and four-fold symmetry instead of full rotational symmetry.
    The interaction strength is fixed to $Na = 1$.
  }
  \label{fig:R2DConfigurations}    
\end{figure}

The starting point of this discussion is the case $\gamma = 0$ of an isolated
system.
For the two-dimensional analysis performed in this section the strength of the
nonlinearity is fixed to $Na = 1$.
To reveal the multiple vortex ground states predicted by Butts et
al.~\cite{Butts99a} various vortex configurations were created to act as
initial values for the nonlinear root search used in the finite-element method.
Using imaginary time propagation and careful tracing of all branches of
solution found, we identified a total of six different ground states in the
range of the rotation frequency from $\Omega = 0$ to $\Omega =0.94$.
Four of these states are shown in Fig.~\ref{fig:R2DConfigurations}, possessing
one ($v_1$ in (a)), two ($v_2$ in (b),(c)), three ($v_3$ in (d)), or four
($v_4$ in (e),(f)) vortices.
These states, with the exception of the state with one central vortex, $v_1$,
do not have rotational symmetry but instead show a two-, three- and four-fold
symmetry, respectively.
For the sake of completeness we note that in the parameter regime discussed in
this section, an additional ground state with four vortices but twofold
symmetry can be observed for $\Omega = 0.9315$.
Applying higher rotation frequencies, the number of vortices would increase
even further, until hexagonal vortex grids can be observed~\cite{Fetter09a}.

By switching on the $\PT$-symmetric
potential~\eqref{eq:harmonicOscillatorImag2D}, the stationary states $v_1$,
$v_2$, $v_3$ and $v_4$ become subject to a gain and loss of particles.
The imaginary part of the potential, and therefore the particle exchange
rotates concurrently with the wave function.
The $\PT$-symmetric potential at $t=0$ induces a current in $y$-direction.
Real eigenvalues and stable behavior can therefore only be expected if the
particle density at $\gamma = 0$ is symmetric with respect to the reflection $y
\to -y$.
Figure~\ref{fig:R2DConfigurations} shows all possible orientations of the four
states that fulfill this requirement.

\begin{figure} 
  \centering
  \includegraphics{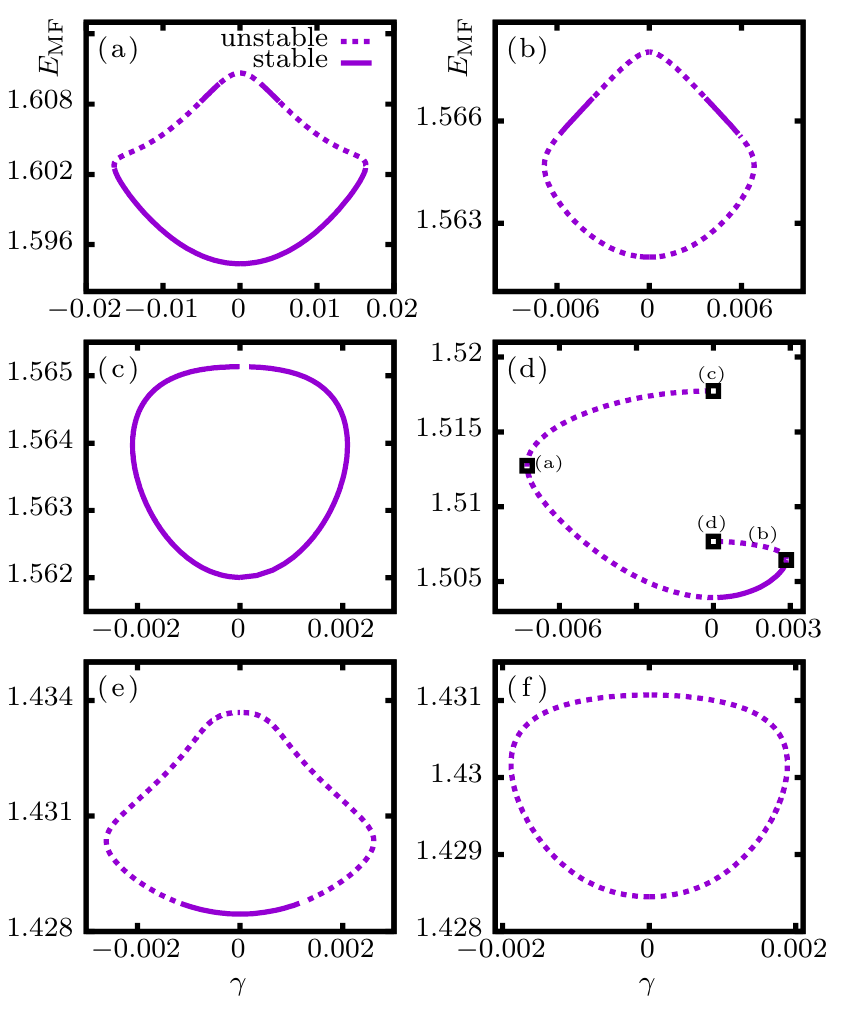}%
  \caption{
    Mean-field energy of the four ground states $v_1$ at $\Omega = 0.85$ (a),
    $v_2$ at $\Omega = 0.88$ (b),(c), $v_3$ at $\Omega = 0.91$ 
    (d) and $v_4$ at $\Omega = 0.94$ (e),(f) for $Na = 1$ as a function of the
    in- and out-coupling strength $\gamma$.
    The orientations are chosen in the same way as presented in
    Fig.~\ref{fig:R2DConfigurations}.
    While both ground states with two vortices on a central axis parallel to
    the current are unstable for any $\gamma \neq 0$ (b),(f), all other
    configurations show stable stationary ground states in some parameter
    regimes.
    All spectra with the exception of the state $v_3$ (d) are even functions of
    $\gamma$.
    Inside the spectra of the state $v_3$ (d), the two tangent bifurcations
    and the involved states at the parameter $\gamma=0$ are marked as a
    reference for Fig.~\ref{fig:R2DWaveFncGamma}.
  }
  \label{fig:R2DSpectra}
\end{figure}

The stability of all states is analyzed via the Bogoliubov-de Gennes equations
in the rotating frame,
\begin{subequations}
\label{eq:BdGE_rotatingFrame}
\begin{align}
  &\left(-\Delta +
      V(\bm{r}) - \mu - \omega + 16 \pi Na \left|\psi_0(\bm{r})\right|^2 -
      \Omega \hat{L}_z
  \right) u(\bm{r})\notag\\
  &+ 8 \pi Na \psi_0^2(\bm{r}) v(\bm{r}) = 0,\\
  &\left(-\Delta +
      V^*(\bm{r}) - \mu^* + \omega + 16 \pi Na \left|\psi_0(\bm{r})\right|^2 +
      \Omega \hat{L}_z
      \right) v(\bm{r})\notag\\
  &+ 8 \pi Na \psi_0^{*2}(\bm{r})u(\bm{r}) = 0.
\end{align}
\end{subequations}
The mean-field energy of the four ground states is shown together with their
dynamical stability in Fig.~\ref{fig:R2DSpectra}.
The most prominent feature of these spectra is that most of them are even
functions of $\gamma$.
This is a consequence of their symmetry which can be seen as follows.
A reflection of both coordinates $x$ and $y$ obviously leaves the
Gross-Pitaevskii equation~\eqref{eq:toSolveRot2D} invariant if the wave
function itself is invariant under this operation.
However, the reflection in $y$-direction inverts the imaginary
potential~\eqref{eq:harmonicOscillatorImag2D} and can be absorbed in the
transformation $\gamma \to -\gamma$.
Therefore, every state that is symmetric in $x$-direction for $\gamma = 0$,
must therefore have an eigenvalue spectrum that is symmetric with respect to
$\gamma = 0$.

\begin{figure} 
  \centering
  \includegraphics{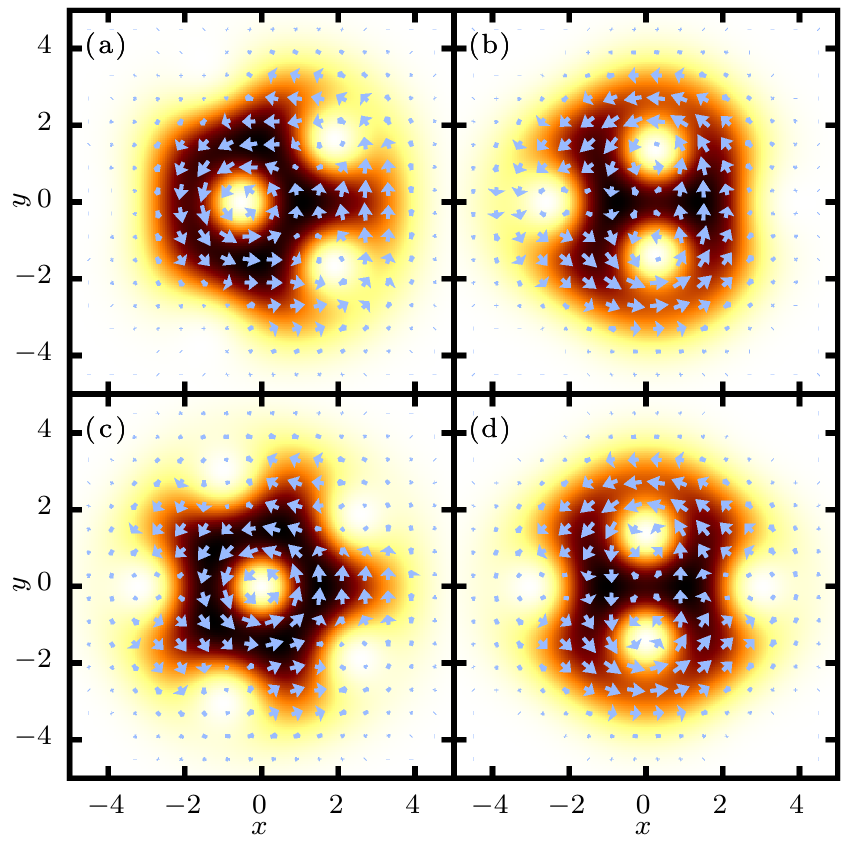}%
  \caption{
    Wave functions marked in the spectra of the state $v_3$ for $\Omega =
    0.91$ and $Na = 1$ shown in Fig.~\ref{fig:R2DSpectra} (d).
    The two states at the bifurcations for negative (a) and positive (b) values
    of $\gamma$ are shown above the respective bifurcation partners (c) and
    (d) taken at $\gamma = 0$.
    The particle density is shown as a color map, in which darker regions
    correspond to higher densities, while the currents are depicted by
    bright blue arrows.
    Their lengths are proportional to the current strength.
  }
  \label{fig:R2DWaveFncGamma}
\end{figure}

All ground states exist up or down to some critical value $\gamma$ where they
coalesce with excited states and vanish (inverse tangent bifurcation).
This is the typical behavior known from $\PT$-symmetric systems.
However, due to the vortices, it is not a priori clear what drives the
necessary current, i.e., how the necessary phase gradient is generated.
To understand the principles of this mechanism the case $v_3$ is studied in
more detail.
Figure~\ref{fig:R2DSpectra}(d) clearly shows that there are two bifurcation
partners and two tangent bifurcations.
The two states at the bifurcations and the bifurcation partners for $\gamma =
0$, as marked in Fig~\ref{fig:R2DSpectra}(d) are shown in
Fig.~\ref{fig:R2DWaveFncGamma}.
For each state not only the square modulus of the wave function is analyzed but
the particle current density in the non-rotating frame is studied in detail
to understand how the currents around the vortices contribute to the net
current enforced by the particle gain and loss.

For negative values of $\gamma$ the net current in the wave function (a) must
run downwards.
This behavior is produced by shifting all three vortex centers to the right
while keeping the overall particle density mostly intact.
This results in a situation where the current in downward direction on the left
side of the wave function is strongly enhanced due to the stronger density.
The exact opposite behavior can be found in the case of positive values of
$\gamma$, where all vortices shift to the left increasing the net current
upwards (b).

This is, however, not the only mechanism that changes the net current of these
vortex states.
A second mechanism can be seen when analyzing, how the excited states shown in
Fig.~\ref{fig:R2DWaveFncGamma}(c) and (d) behave between $\gamma = 0$ and the
bifurcation ((a) and (b)).
In complete contrast to the behavior of the ground state, not all vortices
are shifted equally strong.
Instead, some vortices are shifted more strongly, completely moving out of the
condensate.
This increases the current on this side of their neighboring vortices producing
an effective current in up- or downwards direction.

For the sake of brevity, we omit a detailed discussion of the other five
configurations.
A in-depth study shows that these two effects capture almost every current
production found in the discussed ground states.
In some cases, like the central vortex state $v_1$, the net current is produced
by weakening the undesired current by a new vortex.
In other cases, such as the configuration in
Fig.~\ref{fig:R2DConfigurations}(c) of two vortices lying on the symmetry axis,
both mechanisms are found nearly canceling each other out.
In this case the wave function must undergo a serious transformation even
though the net currents are very weak.
The bifurcation partners, analyzed at $\gamma = 0$, cannot only have a
different number of vortices but may also be asymmetric with respect to the
reflection $x \to -x$.

The results show some similarities to the results of~\cite{Achilleos12a}, where
$\PT$-symmetric bifurcation partners of highly excited nonlinear vortex states
are always states with one vortex more or less.
However, in contrast to that work and due to the rotation contribution in our
calculations, the vortex configurations are much more stable and vortices only
emerge from the wave functions border and not out of a nodal plane.

\section{Three-dimensional system}
\label{sec:3D}
Until now, the current always ran perpendicular to the vortex lines.
This is a consequence of the restriction to two dimensions.
In this section, a net current in direction of the vortices is considered.
If the gain and loss of particles is homogeneous in the
$x$-$y$-plane and varies only in $z$-direction, the imaginary part of the
potential does not even need to rotate.
In contrast to the two-dimensional case, the trapping in $z$-direction is not
taken to be tight but equal to the radial trapping.
With this condition, $\omega_r = \omega_z$, an isotropic trapping potential is
obtained,
\begin{equation}
  V(r) = \frac{1}{4} r^2.
  \label{eq:harmonicOscillatorRPotReal}
\end{equation}
The imaginary potential
\begin{equation}
  V_I(z) = - \rmi \gamma\, \text{sign}(z)
  \label{eq:harmonicOscillatorImag3D}
\end{equation}
with positive values of $\gamma$ implements a gain of particles below the
$x$-$y$-plane and a loss of particles above.
The imaginary part of the potential is again constant in each of these regions
to ensure that an equal amount of particles is coupled in and out of the system
for any shape of the wave function, of which the particle density is at least
symmetric under the transformation $z \to -z$.

To gain access to numerical results, the finite-element method must be provided
with initial values for its root search.
As long as one is interested in the ground state of the system, excitations in
the $z$-direction can be excluded from this search.
A good approximation to the three-dimensional wave function can therefore be
found using a product state of the two-dimensional solution in $x$- and
$y$-directions and a Gaussian ground state in $z$-direction.

However, to reliably postulate initial values for the ground state the two-
and three-dimensional systems must be comparable to start with.
Since the trapping potential in $z$-direction is much weaker than in the
two-dimensional case, the modulus square of the mean-field wave function is
smaller.
To counterbalance this effect the particle number is increased to $Na = 5$.
The Gross-Pitaevskii equation then reads
\begin{align}
  \label{eq:toSolveRot3D}
  \mu& \psi(\bm{r}) =\\
  &\left(-\Delta + \frac{1}{4} r^2 - \rmi \gamma
  \,\text{sign}(z) + 40 \pi |\psi(\bm{r})|^2 - \Omega \hat{L}_z \right)
  \psi(\bm{r}). \notag
\end{align}

\begin{figure} 
  \centering
  \includegraphics{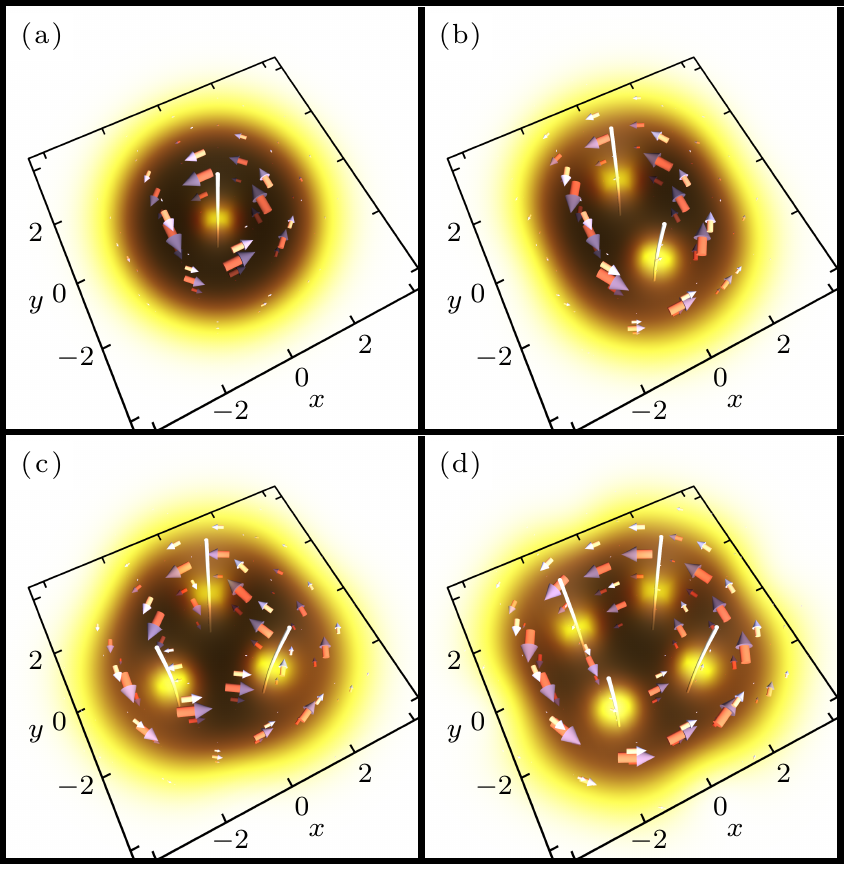}%
  \caption{
    The particle density of the four ground states in three dimensions with one
    to four vortices is shown as a color map, in which darker regions
    correspond to higher densities, while the currents are depicted by
    bright blue-headed arrows.
    The vortex centers are highlighted by white lines.
    For $\Omega = 0.85$ (a) one central vortex exists, while for $\Omega =
    0.87$ (b), $\Omega = 0.9$ (c) and $\Omega = 0.94$ (d) all vortices are
    located off center.
  }
  \label{fig:R3DWaveFunctionGamma0}    
\end{figure}

We start by presenting the rotating ground states for four different rotation
frequencies in Fig.~\ref{fig:R3DWaveFunctionGamma0}.
As in Sec.~\ref{sec:2D} all wave functions are shown in the non-rotating
laboratory system at time $t=0$.
Therefore the figure reveals an overall circular current of particles that is
consistent with the rotation of the wave function.
The concrete path of a vortex core in $z$-direction is defined by its nodal
line.
To permit a clear view on this path, the nodal line is highlighted by
white lines.
Special attention should be given to the state with three vortices in
Fig.~\ref{fig:R3DWaveFunctionGamma0}(c).
Around $z = 0$ the vortex lines are bent inwards towards the rotation center.
This effect, even though most distinct at (c), is present in all three wave
functions with vortices outside the rotation center.
This can be understood since the Magnus force, which keeps the vortices on
their circular track during the rotation, is stronger for regions of higher
particle density, i.e., the density gradient and therefore the Magnus
force~\cite{Anderson00a} is increased.

\begin{figure} 
  \centering
  \includegraphics{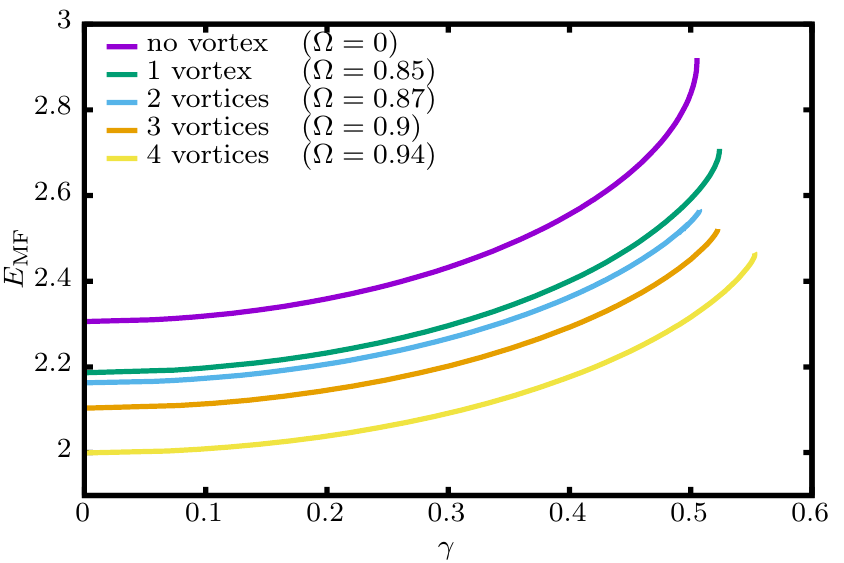}%
  \caption{
    Mean-field energy of the ground states with zero to four vortices.
    The in- and out-coupling parameter $\gamma$ is increased until the
    mean-field energy undergoes a tangent bifurcation, at which the states
    vanish.
    The bifurcation partners are not shown and correspond to the same states
    with an additional excitation in $z$-direction.
  }
  \label{fig:R3DSpectrumGamma}    
\end{figure}

By increasing the parameter $\gamma$, the in- and out-coupling drives a new
current in the system; particles now have to be transported upwards parallel to
the vortices.
In case of the unrotated ground state and the four states discussed in
Fig.~\ref{fig:R3DWaveFunctionGamma0} this leads to an increase of the
mean-field energy, as shown in Fig.~\ref{fig:R3DSpectrumGamma}.
The first eye-catching result of this analysis is that all these ground states
behave qualitatively the same.
In fact, the bifurcation scenario resembles the behavior of the double-well
system or the harmonic oscillator studied in previous
investigations~\cite{Cartarius12b, Dast13a}.
All states break the $\PT$ symmetry shortly after $\gamma = 0.5$ in a typical
tangent bifurcation.
The solutions of the Bogoliubov-de Gennes
equations~\eqref{eq:BdGE_rotatingFrame} show that they are stable up to this
point.

Since the rotation, controlled by the parameter $\Omega$, changes the wave
function considerably, not only by increasing or decreasing the number of
vortices but also by broadening the wave function, this similarity is quite
surprising.
It indicates that the $x$-$y$-plane and the $z$-dimension are only weakly
coupled, even though the nonlinearity already provides a major contribution to
the planar solutions.

\begin{figure} 
  \centering
  \includegraphics{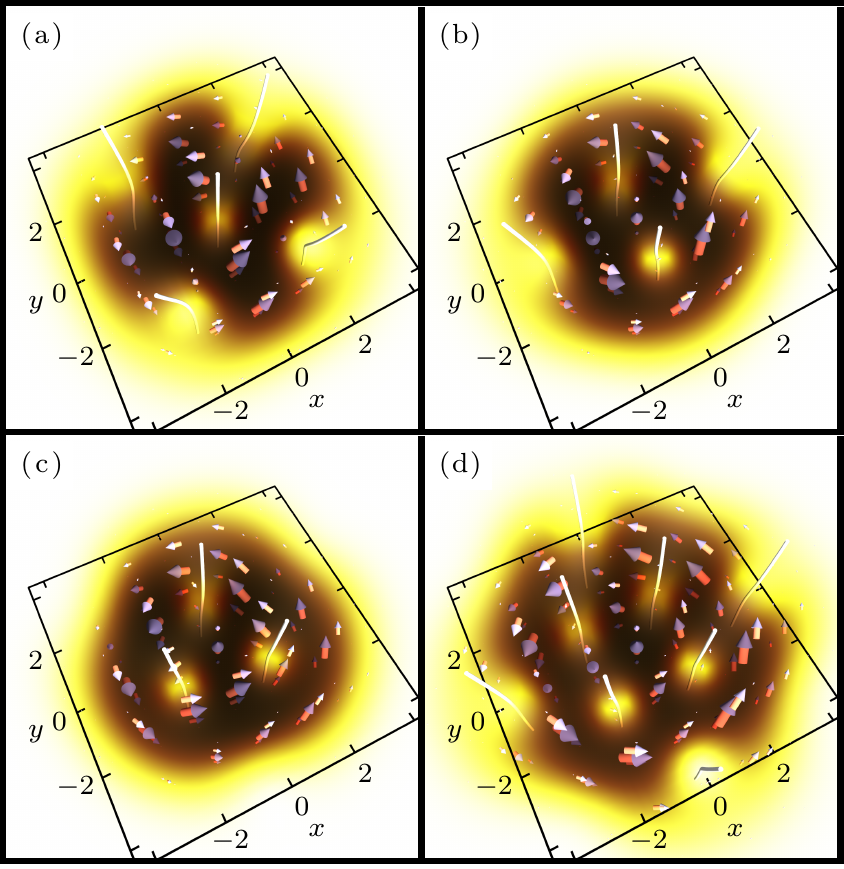}%
  \caption{
    The four ground states with one to four vortices after evolving to their
    maximum $\gamma$.
    The particle density is shown as a color map, in which darker regions
    correspond to higher densities, while the currents are depicted by
    bright blue-headed arrows.
    The vortex centers are highlighted by white lines.
    Additional vortices are added to the ground state at $\Omega = 0.85$ and
    $\gamma = 0.52$ (a), $\Omega = 0.87,\, \gamma = 0.5$ (b) and $\Omega =
    0.94,\, \gamma = 0.54$ (d).
    The three vortex state $\Omega = 0.9,\, \gamma = 0.51$ (c) is mainly
    unchanged.
  }
  \label{fig:R3DWaveFunctionGammaMax}    
\end{figure}

At the bifurcation point the wave functions support the strongest possible
current upwards.
Figure~\ref{fig:R3DWaveFunctionGammaMax} shows these wave functions for the
same states as in Fig.~\ref{fig:R3DWaveFunctionGamma0}.
Two important effects are visible in these wave functions with maximum-current:
Firstly, the number and position of the vortices in the $x$-$y$-plane are
changing.
This is easy to see in Fig.~\ref{fig:R3DWaveFunctionGammaMax}(d).
Not only have four new vortices entered the picture, the original vortices are
pushed much tighter together.
The new vortices have also increased the size of the wave function.
This effect would be expected if either the rotation frequency is increased, or
a the interaction strenghened.
Due to the $\PT$-symmetric current in $z$-direction this component of the wave
function cannot be chosen exactly symmetric, i.e., it does not take the shape
of a Gaussian.
Instead, an antisymmetric contribution is needed, considerably reducing the
modulus square at $z=0$.
The particles are then forced to the top and bottom of the trap, increasing the
particle density and the effective strength of the interaction at these areas.

Secondly, the previously discussed bending to the center of the trap is not the
only deformation of the vortex lines.
Following the direction of the $\PT$-symmetric current upwards, the vortex
lines are screwed in clockwise direction, i.e., against the direction of the
frame's rotation.
To quantify this screwing, each vortex must be parametrized by the coordinate
$z$ in cylindrical coordinates $r(z), \phi(z)$.
The vortex screwing is then purely described by the function $\phi(z)$, which
is antisymmetric with respect to the $x$-$y$-plane;
the differential $\rmd\phi/\rmd z |_{z=0}$ defines a screwing strength.
\begin{figure} 
  \centering
  \includegraphics{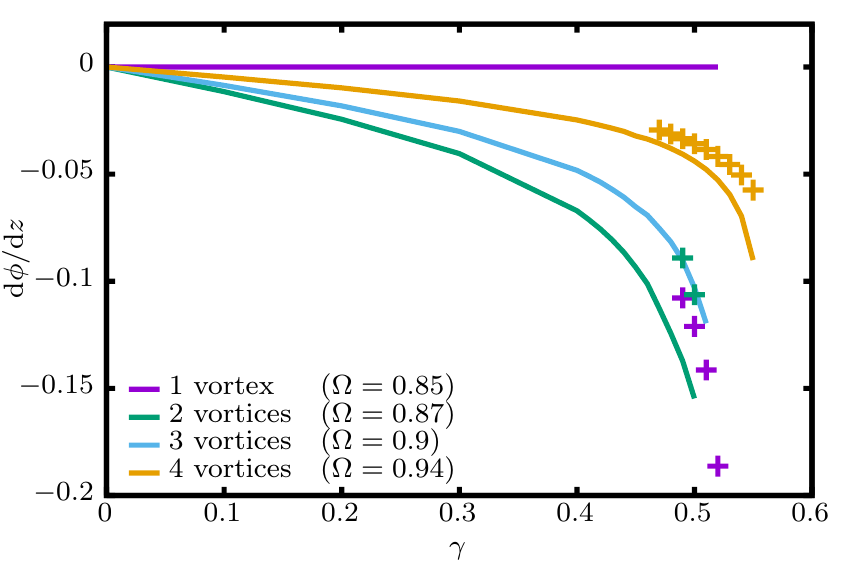}%
  \caption{
    The screwing strength given as $\rmd\phi / \rmd z|_{z=0}$ of the
    parametrized vortices, as a function of $\gamma$ and for different rotation
    frequencies $\Omega$.
    The original vortices existing from $\gamma = 0$ upwards are shown as solid
    lines while the new vortices arising for larger parameters $\gamma$ are
    depicted as crosses.
    Note, that all vortices of the same type and from the same wave function
    are shifted equally.
  }
  \label{fig:VortexAnalyse}    
\end{figure}
This value is shown in Fig.~\ref{fig:VortexAnalyse} as a function of $\gamma$.

The shape of all these functions are very similar.
In fact, only the maximum reachable $\gamma$ and the overall slope differ.
The different maximum parameters $\gamma$ are an obvious consequence of the
different positions of the tangent bifurcations at which the ground states
vanish.
The different slopes are best visible for small parameters $\gamma$.
In this regime, two qualitative dependencies are visible:
Firstly, the central vortex is not bent at all.
Secondly, in a stationary state with $n$ non-center vortices they are screwed
$m/n$ times as strong as in the case of $m$ vortices.

This fact indicates that the vortex screwing supports the upward current in
the system and each vortex makes an equal contribution.
The strongest screwed vortices are therefore found in the two-vortex case.
For stronger gain and loss new non-center vortices arise, depicted as crosses
in Fig.~\ref{fig:VortexAnalyse}.
In case of the central vortex state, these new vortices are the only screwed
vortices in the wave function.
However, in the other two cases the additional vortices are screwed less than
the original ones.

\section{Conclusion}
\label{sec:Conclusion}
We studied a rotating Bose-Einstein condensate with a $\PT$-symmetric potential
describing particle in- and out-coupling.
Initially, the particle transport through the ground state with multiple
vortices was studied for the two-dimensional case where the net currents run
perpendicular to the rotation axis, i.e., the vortex lines.
Not only remain most of the states stable, at least for weak currents, but the
behavior of the states producing such currents shows an interesting behavior.
Either new vortices enter the condensate from the border, weakening parts of
the circular currents around existing vortices, or the existing vortices move
through the non-homogeneous particle distribution to modify the effective
currents.

Both effects obviously arise from the finite size of the condensate, i.e., the
drop of the density at its borders.
Therefore, it would be worthwhile to study whether the transport phenomena
described by $\PT$-symmetric in- and out-coupling is indeed dominated by border
currents and border effects of the condensate.
The next step would be the study of a larger condensate with constant trapping
potential and a bigger vortex grid.

In the next part we added the third dimension to study currents parallel to
the rotation axis, i.e., in direction of the vortex lines.
In the two-dimensional study weak perpendicular currents sufficed to modify
the vortex structure substantially.
This changes drastically in the three-dimensional case, in which even strong
parallel currents do not break the $\PT$-symmetry.
However, the trajectory of the vortex lines in the stable ground states
changes.
Stronger particle in- and out-coupling strengths lead to a screwing of the
vortex lines against the direction of the rotation.

There are various starting points for future studies.
While an additional analysis of the stationary system could provide insight into
the physical process that leads to the screwing, dynamical calculations will
allow us to study how the system behaves when turning on the particle
transport.
Since the screwed states are stable, the ground state without particle gain and
loss can be considered being a small perturbation to the screwed case, and we
expect that the current excites oscillations of the vortex lines.
For example, it should be possible to study the process in the framework of the
dynamics of single vortices in superfluids~\cite{Kevrekidis03a,
Frantzeskakis02a, Anderson00a} and discuss, whether the screwing can be
understood as a result of the Magnus force.


\begin{thebibliography}{57}%
\makeatletter
\providecommand \@ifxundefined [1]{%
 \@ifx{#1\undefined}
}%
\providecommand \@ifnum [1]{%
 \ifnum #1\expandafter \@firstoftwo
 \else \expandafter \@secondoftwo
 \fi
}%
\providecommand \@ifx [1]{%
 \ifx #1\expandafter \@firstoftwo
 \else \expandafter \@secondoftwo
 \fi
}%
\providecommand \natexlab [1]{#1}%
\providecommand \enquote  [1]{``#1''}%
\providecommand \bibnamefont  [1]{#1}%
\providecommand \bibfnamefont [1]{#1}%
\providecommand \citenamefont [1]{#1}%
\providecommand \href@noop [0]{\@secondoftwo}%
\providecommand \href [0]{\begingroup \@sanitize@url \@href}%
\providecommand \@href[1]{\@@startlink{#1}\@@href}%
\providecommand \@@href[1]{\endgroup#1\@@endlink}%
\providecommand \@sanitize@url [0]{\catcode `\\12\catcode `\$12\catcode
  `\&12\catcode `\#12\catcode `\^12\catcode `\_12\catcode `\%12\relax}%
\providecommand \@@startlink[1]{}%
\providecommand \@@endlink[0]{}%
\providecommand \url  [0]{\begingroup\@sanitize@url \@url }%
\providecommand \@url [1]{\endgroup\@href {#1}{\urlprefix }}%
\providecommand \urlprefix  [0]{URL }%
\providecommand \Eprint [0]{\href }%
\providecommand \doibase [0]{http://dx.doi.org/}%
\providecommand \selectlanguage [0]{\@gobble}%
\providecommand \bibinfo  [0]{\@secondoftwo}%
\providecommand \bibfield  [0]{\@secondoftwo}%
\providecommand \translation [1]{[#1]}%
\providecommand \BibitemOpen [0]{}%
\providecommand \bibitemStop [0]{}%
\providecommand \bibitemNoStop [0]{.\EOS\space}%
\providecommand \EOS [0]{\spacefactor3000\relax}%
\providecommand \BibitemShut  [1]{\csname bibitem#1\endcsname}%
\let\auto@bib@innerbib\@empty
\bibitem [{\citenamefont {Fetter}(2009)}]{Fetter09a}%
  \BibitemOpen
  \bibfield  {author} {\bibinfo {author} {\bibfnamefont {A.~L.}\ \bibnamefont
  {Fetter}},\ }\href {\doibase 10.1103/RevModPhys.81.647} {\bibfield  {journal}
  {\bibinfo  {journal} {Rev. Mod. Phys.}\ }\textbf {\bibinfo {volume} {81}},\
  \bibinfo {pages} {647} (\bibinfo {year} {2009})}\BibitemShut {NoStop}%
\bibitem [{\citenamefont {Feynman}(1955)}]{Feynman55a}%
  \BibitemOpen
  \bibfield  {author} {\bibinfo {author} {\bibfnamefont {R.}~\bibnamefont
  {Feynman}},\ }\href {\doibase 10.1016/S0079-6417(08)60077-3} {\bibfield
  {journal} {\bibinfo  {journal} {Prog. Low Temp. Phys.}\ }\textbf {\bibinfo
  {volume} {1}},\ \bibinfo {pages} {17 } (\bibinfo {year} {1955})}\BibitemShut
  {NoStop}%
\bibitem [{\citenamefont {Abrikosov}(1957)}]{Abrikosov57a}%
  \BibitemOpen
  \bibfield  {author} {\bibinfo {author} {\bibfnamefont {A.~A.}\ \bibnamefont
  {Abrikosov}},\ }\href@noop {} {\bibfield  {journal} {\bibinfo  {journal}
  {Sov. Phys. JETP}\ }\textbf {\bibinfo {volume} {5}},\ \bibinfo {pages} {1174}
  (\bibinfo {year} {1957})}\BibitemShut {NoStop}%
\bibitem [{\citenamefont {Pitaevskii}(1961)}]{Pitaevskii61a}%
  \BibitemOpen
  \bibfield  {author} {\bibinfo {author} {\bibfnamefont {L.~P.}\ \bibnamefont
  {Pitaevskii}},\ }\href
  {http://www.jetp.ac.ru/cgi-bin/e/index/e/13/2/p451?a=list} {\bibfield
  {journal} {\bibinfo  {journal} {Sov. Phys. JETP}\ }\textbf {\bibinfo {volume}
  {13}},\ \bibinfo {pages} {451} (\bibinfo {year} {1961})}\BibitemShut
  {NoStop}%
\bibitem [{\citenamefont {Gross}(1961)}]{Gross61a}%
  \BibitemOpen
  \bibfield  {author} {\bibinfo {author} {\bibfnamefont {E.~P.}\ \bibnamefont
  {Gross}},\ }\href {\doibase 10.1007/BF02731494} {\bibfield  {journal}
  {\bibinfo  {journal} {Nuovo Cimento}\ }\textbf {\bibinfo {volume} {20}},\
  \bibinfo {pages} {454} (\bibinfo {year} {1961})}\BibitemShut {NoStop}%
\bibitem [{\citenamefont {Davis}\ \emph {et~al.}(1995)\citenamefont {Davis},
  \citenamefont {Mewes}, \citenamefont {Andrews}, \citenamefont {van Druten},
  \citenamefont {Durfee}, \citenamefont {Kurn},\ and\ \citenamefont
  {Ketterle}}]{Davis95a}%
  \BibitemOpen
  \bibfield  {author} {\bibinfo {author} {\bibfnamefont {K.~B.}\ \bibnamefont
  {Davis}}, \bibinfo {author} {\bibfnamefont {M.-O.}\ \bibnamefont {Mewes}},
  \bibinfo {author} {\bibfnamefont {M.~R.}\ \bibnamefont {Andrews}}, \bibinfo
  {author} {\bibfnamefont {N.~J.}\ \bibnamefont {van Druten}}, \bibinfo
  {author} {\bibfnamefont {D.~S.}\ \bibnamefont {Durfee}}, \bibinfo {author}
  {\bibfnamefont {D.~M.}\ \bibnamefont {Kurn}}, \ and\ \bibinfo {author}
  {\bibfnamefont {W.}~\bibnamefont {Ketterle}},\ }\href {\doibase
  10.1103/PhysRevLett.75.3969} {\bibfield  {journal} {\bibinfo  {journal}
  {Phys. Rev. Lett.}\ }\textbf {\bibinfo {volume} {75}},\ \bibinfo {pages}
  {3969} (\bibinfo {year} {1995})}\BibitemShut {NoStop}%
\bibitem [{\citenamefont {Anderson}\ \emph {et~al.}(1995)\citenamefont
  {Anderson}, \citenamefont {Ensher}, \citenamefont {Matthews}, \citenamefont
  {Wieman},\ and\ \citenamefont {Cornell}}]{Anderson95a}%
  \BibitemOpen
  \bibfield  {author} {\bibinfo {author} {\bibfnamefont {M.~H.}\ \bibnamefont
  {Anderson}}, \bibinfo {author} {\bibfnamefont {J.~R.}\ \bibnamefont
  {Ensher}}, \bibinfo {author} {\bibfnamefont {M.~R.}\ \bibnamefont
  {Matthews}}, \bibinfo {author} {\bibfnamefont {C.~E.}\ \bibnamefont
  {Wieman}}, \ and\ \bibinfo {author} {\bibfnamefont {E.~A.}\ \bibnamefont
  {Cornell}},\ }\href {\doibase 10.1126/science.269.5221.198} {\bibfield
  {journal} {\bibinfo  {journal} {Science}\ }\textbf {\bibinfo {volume}
  {269}},\ \bibinfo {pages} {198} (\bibinfo {year} {1995})}\BibitemShut
  {NoStop}%
\bibitem [{\citenamefont {Butts}\ and\ \citenamefont
  {Rokhsar}(1999)}]{Butts99a}%
  \BibitemOpen
  \bibfield  {author} {\bibinfo {author} {\bibfnamefont {D.~A.}\ \bibnamefont
  {Butts}}\ and\ \bibinfo {author} {\bibfnamefont {D.~S.}\ \bibnamefont
  {Rokhsar}},\ }\href {\doibase 10.1038/16865} {\bibfield  {journal} {\bibinfo
  {journal} {Nature}\ }\textbf {\bibinfo {volume} {397}},\ \bibinfo {pages}
  {327} (\bibinfo {year} {1999})}\BibitemShut {NoStop}%
\bibitem [{\citenamefont {Madison}\ \emph {et~al.}(2000)\citenamefont
  {Madison}, \citenamefont {Chevy}, \citenamefont {Wohlleben},\ and\
  \citenamefont {Dalibard}}]{Madison00a}%
  \BibitemOpen
  \bibfield  {author} {\bibinfo {author} {\bibfnamefont {K.~W.}\ \bibnamefont
  {Madison}}, \bibinfo {author} {\bibfnamefont {F.}~\bibnamefont {Chevy}},
  \bibinfo {author} {\bibfnamefont {W.}~\bibnamefont {Wohlleben}}, \ and\
  \bibinfo {author} {\bibfnamefont {J.}~\bibnamefont {Dalibard}},\ }\href
  {\doibase 10.1103/PhysRevLett.84.806} {\bibfield  {journal} {\bibinfo
  {journal} {Phys. Rev. Lett.}\ }\textbf {\bibinfo {volume} {84}},\ \bibinfo
  {pages} {806} (\bibinfo {year} {2000})}\BibitemShut {NoStop}%
\bibitem [{\citenamefont {Bao}(2007)}]{Bao07a}%
  \BibitemOpen
  \bibfield  {author} {\bibinfo {author} {\bibfnamefont {W.}~\bibnamefont
  {Bao}},\ }\enquote {\bibinfo {title} {{Ground states and dynamics of rotating
  Bose-Einstein condensates}},}\ in\ \href {\doibase
  10.1007/978-0-8176-4554-0_10} {\emph {\bibinfo {booktitle} {{Transport
  Phenomena and Kinetic Theory: Applications to Gases, Semiconductors, Photons,
  and Biological Systems}}}},\ \bibinfo {editor} {edited by\ \bibinfo {editor}
  {\bibfnamefont {C.}~\bibnamefont {Cercignani}}\ and\ \bibinfo {editor}
  {\bibfnamefont {E.}~\bibnamefont {Gabetta}}}\ (\bibinfo  {publisher}
  {Birkh\"auser Boston},\ \bibinfo {address} {Boston, MA},\ \bibinfo {year}
  {2007})\ pp.\ \bibinfo {pages} {215--255}\BibitemShut {NoStop}%
\bibitem [{\citenamefont {Choi}\ and\ \citenamefont {Niu}(1999)}]{Choi99a}%
  \BibitemOpen
  \bibfield  {author} {\bibinfo {author} {\bibfnamefont {D.-I.}\ \bibnamefont
  {Choi}}\ and\ \bibinfo {author} {\bibfnamefont {Q.}~\bibnamefont {Niu}},\
  }\href {\doibase 10.1103/PhysRevLett.82.2022} {\bibfield  {journal} {\bibinfo
   {journal} {Phys. Rev. Lett.}\ }\textbf {\bibinfo {volume} {82}},\ \bibinfo
  {pages} {2022} (\bibinfo {year} {1999})}\BibitemShut {NoStop}%
\bibitem [{\citenamefont {Cl\'ement}\ \emph {et~al.}(2005)\citenamefont
  {Cl\'ement}, \citenamefont {Var\'on}, \citenamefont {Hugbart}, \citenamefont
  {Retter}, \citenamefont {Bouyer}, \citenamefont {Sanchez-Palencia},
  \citenamefont {Gangardt}, \citenamefont {Shlyapnikov},\ and\ \citenamefont
  {Aspect}}]{Clement05a}%
  \BibitemOpen
  \bibfield  {author} {\bibinfo {author} {\bibfnamefont {D.}~\bibnamefont
  {Cl\'ement}}, \bibinfo {author} {\bibfnamefont {A.~F.}\ \bibnamefont
  {Var\'on}}, \bibinfo {author} {\bibfnamefont {M.}~\bibnamefont {Hugbart}},
  \bibinfo {author} {\bibfnamefont {J.~A.}\ \bibnamefont {Retter}}, \bibinfo
  {author} {\bibfnamefont {P.}~\bibnamefont {Bouyer}}, \bibinfo {author}
  {\bibfnamefont {L.}~\bibnamefont {Sanchez-Palencia}}, \bibinfo {author}
  {\bibfnamefont {D.~M.}\ \bibnamefont {Gangardt}}, \bibinfo {author}
  {\bibfnamefont {G.~V.}\ \bibnamefont {Shlyapnikov}}, \ and\ \bibinfo {author}
  {\bibfnamefont {A.}~\bibnamefont {Aspect}},\ }\href {\doibase
  10.1103/PhysRevLett.95.170409} {\bibfield  {journal} {\bibinfo  {journal}
  {Phys. Rev. Lett.}\ }\textbf {\bibinfo {volume} {95}},\ \bibinfo {pages}
  {170409} (\bibinfo {year} {2005})}\BibitemShut {NoStop}%
\bibitem [{\citenamefont {Putterman}(1974)}]{Putterman74a}%
  \BibitemOpen
  \bibfield  {author} {\bibinfo {author} {\bibfnamefont {S.~J.}\ \bibnamefont
  {Putterman}},\ }\href@noop {} {\emph {\bibinfo {title} {Superfluid
  hydrodynamics}}}\ (\bibinfo  {publisher} {North-Holland},\ \bibinfo {address}
  {Amsterdam},\ \bibinfo {year} {1974})\BibitemShut {NoStop}%
\bibitem [{\citenamefont {Pethick}\ \emph {et~al.}(1977)\citenamefont
  {Pethick}, \citenamefont {Smith},\ and\ \citenamefont
  {Bhattacharyya}}]{Pethick77a}%
  \BibitemOpen
  \bibfield  {author} {\bibinfo {author} {\bibfnamefont {C.~J.}\ \bibnamefont
  {Pethick}}, \bibinfo {author} {\bibfnamefont {H.}~\bibnamefont {Smith}}, \
  and\ \bibinfo {author} {\bibfnamefont {P.}~\bibnamefont {Bhattacharyya}},\
  }\href {\doibase 10.1103/PhysRevB.15.3384} {\bibfield  {journal} {\bibinfo
  {journal} {Phys. Rev. B}\ }\textbf {\bibinfo {volume} {15}},\ \bibinfo
  {pages} {3384} (\bibinfo {year} {1977})}\BibitemShut {NoStop}%
\bibitem [{\citenamefont {Campbell}\ and\ \citenamefont
  {Evetts}(1972)}]{Campbell72a}%
  \BibitemOpen
  \bibfield  {author} {\bibinfo {author} {\bibfnamefont {A.}~\bibnamefont
  {Campbell}}\ and\ \bibinfo {author} {\bibfnamefont {J.}~\bibnamefont
  {Evetts}},\ }\href {\doibase 10.1080/00018737200101288} {\bibfield  {journal}
  {\bibinfo  {journal} {Adv. Phys.}\ }\textbf {\bibinfo {volume} {21}},\
  \bibinfo {pages} {199} (\bibinfo {year} {1972})}\BibitemShut {NoStop}%
\bibitem [{\citenamefont {Kagan}\ \emph {et~al.}(1998)\citenamefont {Kagan},
  \citenamefont {Muryshev},\ and\ \citenamefont {Shlyapnikov}}]{Kagan98a}%
  \BibitemOpen
  \bibfield  {author} {\bibinfo {author} {\bibfnamefont {Y.}~\bibnamefont
  {Kagan}}, \bibinfo {author} {\bibfnamefont {A.~E.}\ \bibnamefont {Muryshev}},
  \ and\ \bibinfo {author} {\bibfnamefont {G.~V.}\ \bibnamefont
  {Shlyapnikov}},\ }\href {\doibase 10.1103/PhysRevLett.81.933} {\bibfield
  {journal} {\bibinfo  {journal} {Phys. Rev. Lett.}\ }\textbf {\bibinfo
  {volume} {81}},\ \bibinfo {pages} {933} (\bibinfo {year} {1998})}\BibitemShut
  {NoStop}%
\bibitem [{\citenamefont {Moiseyev}(2011)}]{Moiseyev11a}%
  \BibitemOpen
  \bibfield  {author} {\bibinfo {author} {\bibfnamefont {N.}~\bibnamefont
  {Moiseyev}},\ }\href@noop {} {\emph {\bibinfo {title} {{Non-Hermitian Quantum
  Mechanics}}}}\ (\bibinfo  {publisher} {Cambridge University Press},\ \bibinfo
  {address} {Cambridge},\ \bibinfo {year} {2011})\BibitemShut {NoStop}%
\bibitem [{\citenamefont {Schlagheck}\ and\ \citenamefont
  {Paul}(2006)}]{Schlagheck06a}%
  \BibitemOpen
  \bibfield  {author} {\bibinfo {author} {\bibfnamefont {P.}~\bibnamefont
  {Schlagheck}}\ and\ \bibinfo {author} {\bibfnamefont {T.}~\bibnamefont
  {Paul}},\ }\href {\doibase 10.1103/PhysRevA.73.023619} {\bibfield  {journal}
  {\bibinfo  {journal} {Phys. Rev. A}\ }\textbf {\bibinfo {volume} {73}},\
  \bibinfo {pages} {023619} (\bibinfo {year} {2006})}\BibitemShut {NoStop}%
\bibitem [{\citenamefont {Rapedius}\ and\ \citenamefont
  {Korsch}(2009)}]{Rapedius09a}%
  \BibitemOpen
  \bibfield  {author} {\bibinfo {author} {\bibfnamefont {K.}~\bibnamefont
  {Rapedius}}\ and\ \bibinfo {author} {\bibfnamefont {H.~J.}\ \bibnamefont
  {Korsch}},\ }\href {\doibase 10.1088/0953-4075/42/4/044005} {\bibfield
  {journal} {\bibinfo  {journal} {J. Phys. B}\ }\textbf {\bibinfo {volume}
  {42}},\ \bibinfo {pages} {044005} (\bibinfo {year} {2009})}\BibitemShut
  {NoStop}%
\bibitem [{\citenamefont {Rapedius}\ \emph {et~al.}(2010)\citenamefont
  {Rapedius}, \citenamefont {Elsen}, \citenamefont {Witthaut}, \citenamefont
  {Wimberger},\ and\ \citenamefont {Korsch}}]{Rapedius10a}%
  \BibitemOpen
  \bibfield  {author} {\bibinfo {author} {\bibfnamefont {K.}~\bibnamefont
  {Rapedius}}, \bibinfo {author} {\bibfnamefont {C.}~\bibnamefont {Elsen}},
  \bibinfo {author} {\bibfnamefont {D.}~\bibnamefont {Witthaut}}, \bibinfo
  {author} {\bibfnamefont {S.}~\bibnamefont {Wimberger}}, \ and\ \bibinfo
  {author} {\bibfnamefont {H.~J.}\ \bibnamefont {Korsch}},\ }\href {\doibase
  10.1103/PhysRevA.82.063601} {\bibfield  {journal} {\bibinfo  {journal} {Phys.
  Rev. A}\ }\textbf {\bibinfo {volume} {82}},\ \bibinfo {pages} {063601}
  (\bibinfo {year} {2010})}\BibitemShut {NoStop}%
\bibitem [{\citenamefont {Abdullaev}\ \emph {et~al.}(2010)\citenamefont
  {Abdullaev}, \citenamefont {Konotop}, \citenamefont {Salerno},\ and\
  \citenamefont {Yulin}}]{Abdullaev10a}%
  \BibitemOpen
  \bibfield  {author} {\bibinfo {author} {\bibfnamefont {F.~K.}\ \bibnamefont
  {Abdullaev}}, \bibinfo {author} {\bibfnamefont {V.~V.}\ \bibnamefont
  {Konotop}}, \bibinfo {author} {\bibfnamefont {M.}~\bibnamefont {Salerno}}, \
  and\ \bibinfo {author} {\bibfnamefont {A.~V.}\ \bibnamefont {Yulin}},\ }\href
  {\doibase 10.1103/PhysRevE.82.056606} {\bibfield  {journal} {\bibinfo
  {journal} {Phys. Rev. E}\ }\textbf {\bibinfo {volume} {82}},\ \bibinfo
  {pages} {056606} (\bibinfo {year} {2010})}\BibitemShut {NoStop}%
\bibitem [{\citenamefont {Bludov}\ and\ \citenamefont
  {Konotop}(2010)}]{Bludov10a}%
  \BibitemOpen
  \bibfield  {author} {\bibinfo {author} {\bibfnamefont {Y.~V.}\ \bibnamefont
  {Bludov}}\ and\ \bibinfo {author} {\bibfnamefont {V.~V.}\ \bibnamefont
  {Konotop}},\ }\href {\doibase 10.1103/PhysRevA.81.013625} {\bibfield
  {journal} {\bibinfo  {journal} {Phys. Rev. A}\ }\textbf {\bibinfo {volume}
  {81}},\ \bibinfo {pages} {013625} (\bibinfo {year} {2010})}\BibitemShut
  {NoStop}%
\bibitem [{\citenamefont {Witthaut}\ \emph {et~al.}(2011)\citenamefont
  {Witthaut}, \citenamefont {Trimborn}, \citenamefont {Hennig}, \citenamefont
  {Kordas}, \citenamefont {Geisel},\ and\ \citenamefont
  {Wimberger}}]{Witthaut11a}%
  \BibitemOpen
  \bibfield  {author} {\bibinfo {author} {\bibfnamefont {D.}~\bibnamefont
  {Witthaut}}, \bibinfo {author} {\bibfnamefont {F.}~\bibnamefont {Trimborn}},
  \bibinfo {author} {\bibfnamefont {H.}~\bibnamefont {Hennig}}, \bibinfo
  {author} {\bibfnamefont {G.}~\bibnamefont {Kordas}}, \bibinfo {author}
  {\bibfnamefont {T.}~\bibnamefont {Geisel}}, \ and\ \bibinfo {author}
  {\bibfnamefont {S.}~\bibnamefont {Wimberger}},\ }\href {\doibase
  10.1103/PhysRevA.83.063608} {\bibfield  {journal} {\bibinfo  {journal} {Phys.
  Rev. A}\ }\textbf {\bibinfo {volume} {83}},\ \bibinfo {pages} {063608}
  (\bibinfo {year} {2011})}\BibitemShut {NoStop}%
\bibitem [{\citenamefont {Rapedius}(2013)}]{Rapedius13a}%
  \BibitemOpen
  \bibfield  {author} {\bibinfo {author} {\bibfnamefont {K.}~\bibnamefont
  {Rapedius}},\ }\href {\doibase 10.1088/0953-4075/46/12/125301} {\bibfield
  {journal} {\bibinfo  {journal} {J. Phys. B}\ }\textbf {\bibinfo {volume}
  {46}},\ \bibinfo {pages} {125301} (\bibinfo {year} {2013})}\BibitemShut
  {NoStop}%
\bibitem [{\citenamefont {Dast}\ \emph {et~al.}(2014)\citenamefont {Dast},
  \citenamefont {Haag}, \citenamefont {Cartarius},\ and\ \citenamefont
  {Wunner}}]{Dast14a}%
  \BibitemOpen
  \bibfield  {author} {\bibinfo {author} {\bibfnamefont {D.}~\bibnamefont
  {Dast}}, \bibinfo {author} {\bibfnamefont {D.}~\bibnamefont {Haag}}, \bibinfo
  {author} {\bibfnamefont {H.}~\bibnamefont {Cartarius}}, \ and\ \bibinfo
  {author} {\bibfnamefont {G.}~\bibnamefont {Wunner}},\ }\href {\doibase
  10.1103/PhysRevA.90.052120} {\bibfield  {journal} {\bibinfo  {journal} {Phys.
  Rev. A}\ }\textbf {\bibinfo {volume} {90}},\ \bibinfo {pages} {052120}
  (\bibinfo {year} {2014})}\BibitemShut {NoStop}%
\bibitem [{\citenamefont {Gericke}\ \emph {et~al.}(2008)\citenamefont
  {Gericke}, \citenamefont {Wurtz}, \citenamefont {Reitz}, \citenamefont
  {Langen},\ and\ \citenamefont {Ott}}]{Gericke08a}%
  \BibitemOpen
  \bibfield  {author} {\bibinfo {author} {\bibfnamefont {T.}~\bibnamefont
  {Gericke}}, \bibinfo {author} {\bibfnamefont {P.}~\bibnamefont {Wurtz}},
  \bibinfo {author} {\bibfnamefont {D.}~\bibnamefont {Reitz}}, \bibinfo
  {author} {\bibfnamefont {T.}~\bibnamefont {Langen}}, \ and\ \bibinfo {author}
  {\bibfnamefont {H.}~\bibnamefont {Ott}},\ }\href {\doibase 10.1038/nphys1102}
  {\bibfield  {journal} {\bibinfo  {journal} {Nat. Phys.}\ }\textbf {\bibinfo
  {volume} {4}},\ \bibinfo {pages} {949} (\bibinfo {year} {2008})}\BibitemShut
  {NoStop}%
\bibitem [{\citenamefont {Robins}\ \emph {et~al.}(2008)\citenamefont {Robins},
  \citenamefont {Figl}, \citenamefont {Jeppesen}, \citenamefont {Dennis},\ and\
  \citenamefont {Close}}]{Robins08a}%
  \BibitemOpen
  \bibfield  {author} {\bibinfo {author} {\bibfnamefont {N.~P.}\ \bibnamefont
  {Robins}}, \bibinfo {author} {\bibfnamefont {C.}~\bibnamefont {Figl}},
  \bibinfo {author} {\bibfnamefont {M.}~\bibnamefont {Jeppesen}}, \bibinfo
  {author} {\bibfnamefont {G.~R.}\ \bibnamefont {Dennis}}, \ and\ \bibinfo
  {author} {\bibfnamefont {J.~D.}\ \bibnamefont {Close}},\ }\href {\doibase
  10.1038/nphys1027} {\bibfield  {journal} {\bibinfo  {journal} {Nat. Phys.}\
  }\textbf {\bibinfo {volume} {4}},\ \bibinfo {pages} {731} (\bibinfo {year}
  {2008})}\BibitemShut {NoStop}%
\bibitem [{\citenamefont {Bender}\ and\ \citenamefont
  {Boettcher}(1998)}]{Bender98a}%
  \BibitemOpen
  \bibfield  {author} {\bibinfo {author} {\bibfnamefont {C.~M.}\ \bibnamefont
  {Bender}}\ and\ \bibinfo {author} {\bibfnamefont {S.}~\bibnamefont
  {Boettcher}},\ }\href {\doibase 10.1103/PhysRevLett.80.5243} {\bibfield
  {journal} {\bibinfo  {journal} {Phys. Rev. Lett.}\ }\textbf {\bibinfo
  {volume} {80}},\ \bibinfo {pages} {5243} (\bibinfo {year}
  {1998})}\BibitemShut {NoStop}%
\bibitem [{\citenamefont {Bender}\ \emph {et~al.}(2002)\citenamefont {Bender},
  \citenamefont {Brody},\ and\ \citenamefont {Jones}}]{Bender02a}%
  \BibitemOpen
  \bibfield  {author} {\bibinfo {author} {\bibfnamefont {C.~M.}\ \bibnamefont
  {Bender}}, \bibinfo {author} {\bibfnamefont {D.~C.}\ \bibnamefont {Brody}}, \
  and\ \bibinfo {author} {\bibfnamefont {H.~F.}\ \bibnamefont {Jones}},\ }\href
  {\doibase 10.1103/PhysRevLett.89.270401} {\bibfield  {journal} {\bibinfo
  {journal} {Phys. Rev. Lett.}\ }\textbf {\bibinfo {volume} {89}},\ \bibinfo
  {pages} {270401} (\bibinfo {year} {2002})}\BibitemShut {NoStop}%
\bibitem [{\citenamefont {Mostafazadeh}(2008)}]{Mostafazadeh08a}%
  \BibitemOpen
  \bibfield  {author} {\bibinfo {author} {\bibfnamefont {A.}~\bibnamefont
  {Mostafazadeh}},\ }\href {\doibase 10.1088/1751-8113/41/5/055304} {\bibfield
  {journal} {\bibinfo  {journal} {J. Phys. A}\ }\textbf {\bibinfo {volume}
  {41}},\ \bibinfo {pages} {055304} (\bibinfo {year} {2008})}\BibitemShut
  {NoStop}%
\bibitem [{\citenamefont {Mostafazadeh}(2010)}]{Mostafazadeh10a}%
  \BibitemOpen
  \bibfield  {author} {\bibinfo {author} {\bibfnamefont {A.}~\bibnamefont
  {Mostafazadeh}},\ }\href {\doibase 10.1142/S0219887810004816} {\bibfield
  {journal} {\bibinfo  {journal} {Int. J. Geom. Methods Mod. Phys.}\ }\textbf
  {\bibinfo {volume} {07}},\ \bibinfo {pages} {1191} (\bibinfo {year}
  {2010})}\BibitemShut {NoStop}%
\bibitem [{\citenamefont {El-Ganainy}\ \emph {et~al.}(2007)\citenamefont
  {El-Ganainy}, \citenamefont {Makris}, \citenamefont {Christodoulides},\ and\
  \citenamefont {Musslimani}}]{El-Ganainy07a}%
  \BibitemOpen
  \bibfield  {author} {\bibinfo {author} {\bibfnamefont {R.}~\bibnamefont
  {El-Ganainy}}, \bibinfo {author} {\bibfnamefont {K.~G.}\ \bibnamefont
  {Makris}}, \bibinfo {author} {\bibfnamefont {D.~N.}\ \bibnamefont
  {Christodoulides}}, \ and\ \bibinfo {author} {\bibfnamefont {Z.~H.}\
  \bibnamefont {Musslimani}},\ }\href {\doibase 10.1364/OL.32.002632}
  {\bibfield  {journal} {\bibinfo  {journal} {Opt. Lett.}\ }\textbf {\bibinfo
  {volume} {32}},\ \bibinfo {pages} {2632} (\bibinfo {year}
  {2007})}\BibitemShut {NoStop}%
\bibitem [{\citenamefont {Klaiman}\ \emph {et~al.}(2008)\citenamefont
  {Klaiman}, \citenamefont {G\"unther},\ and\ \citenamefont
  {Moiseyev}}]{Klaiman08a}%
  \BibitemOpen
  \bibfield  {author} {\bibinfo {author} {\bibfnamefont {S.}~\bibnamefont
  {Klaiman}}, \bibinfo {author} {\bibfnamefont {U.}~\bibnamefont {G\"unther}},
  \ and\ \bibinfo {author} {\bibfnamefont {N.}~\bibnamefont {Moiseyev}},\
  }\href {\doibase 10.1103/PhysRevLett.101.080402} {\bibfield  {journal}
  {\bibinfo  {journal} {Phys. Rev. Lett.}\ }\textbf {\bibinfo {volume} {101}},\
  \bibinfo {pages} {080402} (\bibinfo {year} {2008})}\BibitemShut {NoStop}%
\bibitem [{\citenamefont {Musslimani}\ \emph {et~al.}(2008)\citenamefont
  {Musslimani}, \citenamefont {Makris}, \citenamefont {El-Ganainy},\ and\
  \citenamefont {Christodoulides}}]{Musslimani08a}%
  \BibitemOpen
  \bibfield  {author} {\bibinfo {author} {\bibfnamefont {Z.~H.}\ \bibnamefont
  {Musslimani}}, \bibinfo {author} {\bibfnamefont {K.~G.}\ \bibnamefont
  {Makris}}, \bibinfo {author} {\bibfnamefont {R.}~\bibnamefont {El-Ganainy}},
  \ and\ \bibinfo {author} {\bibfnamefont {D.~N.}\ \bibnamefont
  {Christodoulides}},\ }\href {\doibase 10.1103/PhysRevLett.100.030402}
  {\bibfield  {journal} {\bibinfo  {journal} {Phys. Rev. Lett.}\ }\textbf
  {\bibinfo {volume} {100}},\ \bibinfo {pages} {030402} (\bibinfo {year}
  {2008})}\BibitemShut {NoStop}%
\bibitem [{\citenamefont {Makris}\ \emph {et~al.}(2008)\citenamefont {Makris},
  \citenamefont {El-Ganainy}, \citenamefont {Christodoulides},\ and\
  \citenamefont {Musslimani}}]{Makris08a}%
  \BibitemOpen
  \bibfield  {author} {\bibinfo {author} {\bibfnamefont {K.~G.}\ \bibnamefont
  {Makris}}, \bibinfo {author} {\bibfnamefont {R.}~\bibnamefont {El-Ganainy}},
  \bibinfo {author} {\bibfnamefont {D.~N.}\ \bibnamefont {Christodoulides}}, \
  and\ \bibinfo {author} {\bibfnamefont {Z.~H.}\ \bibnamefont {Musslimani}},\
  }\href {\doibase 10.1103/PhysRevA.81.063807} {\bibfield  {journal} {\bibinfo
  {journal} {Phys. Rev. Lett.}\ }\textbf {\bibinfo {volume} {100}},\ \bibinfo
  {pages} {103904} (\bibinfo {year} {2008})}\BibitemShut {NoStop}%
\bibitem [{\citenamefont {Makris}\ \emph {et~al.}(2010)\citenamefont {Makris},
  \citenamefont {El-Ganainy}, \citenamefont {Christodoulides},\ and\
  \citenamefont {Musslimani}}]{Makris10a}%
  \BibitemOpen
  \bibfield  {author} {\bibinfo {author} {\bibfnamefont {K.~G.}\ \bibnamefont
  {Makris}}, \bibinfo {author} {\bibfnamefont {R.}~\bibnamefont {El-Ganainy}},
  \bibinfo {author} {\bibfnamefont {D.~N.}\ \bibnamefont {Christodoulides}}, \
  and\ \bibinfo {author} {\bibfnamefont {Z.~H.}\ \bibnamefont {Musslimani}},\
  }\href {\doibase 10.1103/PhysRevA.81.063807} {\bibfield  {journal} {\bibinfo
  {journal} {Phys. Rev. A}\ }\textbf {\bibinfo {volume} {81}},\ \bibinfo
  {pages} {063807} (\bibinfo {year} {2010})}\BibitemShut {NoStop}%
\bibitem [{\citenamefont {Guo}\ \emph {et~al.}(2009)\citenamefont {Guo},
  \citenamefont {Salamo}, \citenamefont {Duchesne}, \citenamefont {Morandotti},
  \citenamefont {Volatier-Ravat}, \citenamefont {Aimez}, \citenamefont
  {Siviloglou},\ and\ \citenamefont {Christodoulides}}]{Guo09a}%
  \BibitemOpen
  \bibfield  {author} {\bibinfo {author} {\bibfnamefont {A.}~\bibnamefont
  {Guo}}, \bibinfo {author} {\bibfnamefont {G.~J.}\ \bibnamefont {Salamo}},
  \bibinfo {author} {\bibfnamefont {D.}~\bibnamefont {Duchesne}}, \bibinfo
  {author} {\bibfnamefont {R.}~\bibnamefont {Morandotti}}, \bibinfo {author}
  {\bibfnamefont {M.}~\bibnamefont {Volatier-Ravat}}, \bibinfo {author}
  {\bibfnamefont {V.}~\bibnamefont {Aimez}}, \bibinfo {author} {\bibfnamefont
  {G.~A.}\ \bibnamefont {Siviloglou}}, \ and\ \bibinfo {author} {\bibfnamefont
  {D.~N.}\ \bibnamefont {Christodoulides}},\ }\href {\doibase
  10.1103/PhysRevLett.103.093902} {\bibfield  {journal} {\bibinfo  {journal}
  {Phys. Rev. Lett.}\ }\textbf {\bibinfo {volume} {103}},\ \bibinfo {pages}
  {093902} (\bibinfo {year} {2009})}\BibitemShut {NoStop}%
\bibitem [{\citenamefont {R{\"{u}}ter}\ \emph {et~al.}(2010)\citenamefont
  {R{\"{u}}ter}, \citenamefont {Makris}, \citenamefont {El-Ganainy},
  \citenamefont {Christodoulides}, \citenamefont {Segev},\ and\ \citenamefont
  {Kip}}]{Ruter10a}%
  \BibitemOpen
  \bibfield  {author} {\bibinfo {author} {\bibfnamefont {C.~E.}\ \bibnamefont
  {R{\"{u}}ter}}, \bibinfo {author} {\bibfnamefont {K.~G.}\ \bibnamefont
  {Makris}}, \bibinfo {author} {\bibfnamefont {R.}~\bibnamefont {El-Ganainy}},
  \bibinfo {author} {\bibfnamefont {D.~N.}\ \bibnamefont {Christodoulides}},
  \bibinfo {author} {\bibfnamefont {M.}~\bibnamefont {Segev}}, \ and\ \bibinfo
  {author} {\bibfnamefont {D.}~\bibnamefont {Kip}},\ }\href {\doibase
  10.1038/nphys1515} {\bibfield  {journal} {\bibinfo  {journal} {Nat. Phys.}\
  }\textbf {\bibinfo {volume} {6}},\ \bibinfo {pages} {192} (\bibinfo {year}
  {2010})}\BibitemShut {NoStop}%
\bibitem [{\citenamefont {Peng}\ \emph {et~al.}(2014)\citenamefont {Peng},
  \citenamefont {\"{O}zdemir}, \citenamefont {Lei}, \citenamefont {Monifi},
  \citenamefont {Gianfreda}, \citenamefont {Long}, \citenamefont {Fan},
  \citenamefont {Nori}, \citenamefont {Bender},\ and\ \citenamefont
  {Yang}}]{Peng14b}%
  \BibitemOpen
  \bibfield  {author} {\bibinfo {author} {\bibfnamefont {B.}~\bibnamefont
  {Peng}}, \bibinfo {author} {\bibfnamefont {{\c{S}}.~K.}\ \bibnamefont
  {\"{O}zdemir}}, \bibinfo {author} {\bibfnamefont {F.}~\bibnamefont {Lei}},
  \bibinfo {author} {\bibfnamefont {F.}~\bibnamefont {Monifi}}, \bibinfo
  {author} {\bibfnamefont {M.}~\bibnamefont {Gianfreda}}, \bibinfo {author}
  {\bibfnamefont {G.~L.}\ \bibnamefont {Long}}, \bibinfo {author}
  {\bibfnamefont {S.}~\bibnamefont {Fan}}, \bibinfo {author} {\bibfnamefont
  {F.}~\bibnamefont {Nori}}, \bibinfo {author} {\bibfnamefont {C.~M.}\
  \bibnamefont {Bender}}, \ and\ \bibinfo {author} {\bibfnamefont
  {L.}~\bibnamefont {Yang}},\ }\href {\doibase 10.1038/nphys2927} {\bibfield
  {journal} {\bibinfo  {journal} {Nat. Phys.}\ }\textbf {\bibinfo {volume}
  {10}},\ \bibinfo {pages} {394} (\bibinfo {year} {2014})}\BibitemShut
  {NoStop}%
\bibitem [{\citenamefont {Regensburger}\ \emph {et~al.}(2012)\citenamefont
  {Regensburger}, \citenamefont {Bersch}, \citenamefont {Miri}, \citenamefont
  {Onishchukov}, \citenamefont {Christodoulides},\ and\ \citenamefont
  {Peschel}}]{Regensburger12a}%
  \BibitemOpen
  \bibfield  {author} {\bibinfo {author} {\bibfnamefont {A.}~\bibnamefont
  {Regensburger}}, \bibinfo {author} {\bibfnamefont {C.}~\bibnamefont
  {Bersch}}, \bibinfo {author} {\bibfnamefont {M.-A.}\ \bibnamefont {Miri}},
  \bibinfo {author} {\bibfnamefont {G.}~\bibnamefont {Onishchukov}}, \bibinfo
  {author} {\bibfnamefont {D.~N.}\ \bibnamefont {Christodoulides}}, \ and\
  \bibinfo {author} {\bibfnamefont {U.}~\bibnamefont {Peschel}},\ }\href
  {\doibase 10.1038/nature11298} {\bibfield  {journal} {\bibinfo  {journal}
  {Nature}\ }\textbf {\bibinfo {volume} {488}},\ \bibinfo {pages} {167}
  (\bibinfo {year} {2012})}\BibitemShut {NoStop}%
\bibitem [{\citenamefont {Schindler}\ \emph {et~al.}(2011)\citenamefont
  {Schindler}, \citenamefont {Li}, \citenamefont {Zheng}, \citenamefont
  {Ellis},\ and\ \citenamefont {Kottos}}]{Schindler11a}%
  \BibitemOpen
  \bibfield  {author} {\bibinfo {author} {\bibfnamefont {J.}~\bibnamefont
  {Schindler}}, \bibinfo {author} {\bibfnamefont {A.}~\bibnamefont {Li}},
  \bibinfo {author} {\bibfnamefont {M.~C.}\ \bibnamefont {Zheng}}, \bibinfo
  {author} {\bibfnamefont {F.~M.}\ \bibnamefont {Ellis}}, \ and\ \bibinfo
  {author} {\bibfnamefont {T.}~\bibnamefont {Kottos}},\ }\href {\doibase
  10.1103/PhysRevA.84.040101} {\bibfield  {journal} {\bibinfo  {journal} {Phys.
  Rev. A}\ }\textbf {\bibinfo {volume} {84}},\ \bibinfo {pages} {040101}
  (\bibinfo {year} {2011})}\BibitemShut {NoStop}%
\bibitem [{\citenamefont {Bender}\ \emph {et~al.}(2013)\citenamefont {Bender},
  \citenamefont {Factor}, \citenamefont {Bodyfelt}, \citenamefont {Ramezani},
  \citenamefont {Christodoulides}, \citenamefont {Ellis},\ and\ \citenamefont
  {Kottos}}]{Bender13a}%
  \BibitemOpen
  \bibfield  {author} {\bibinfo {author} {\bibfnamefont {N.}~\bibnamefont
  {Bender}}, \bibinfo {author} {\bibfnamefont {S.}~\bibnamefont {Factor}},
  \bibinfo {author} {\bibfnamefont {J.~D.}\ \bibnamefont {Bodyfelt}}, \bibinfo
  {author} {\bibfnamefont {H.}~\bibnamefont {Ramezani}}, \bibinfo {author}
  {\bibfnamefont {D.~N.}\ \bibnamefont {Christodoulides}}, \bibinfo {author}
  {\bibfnamefont {F.~M.}\ \bibnamefont {Ellis}}, \ and\ \bibinfo {author}
  {\bibfnamefont {T.}~\bibnamefont {Kottos}},\ }\href {\doibase
  10.1103/PhysRevLett.110.234101} {\bibfield  {journal} {\bibinfo  {journal}
  {Phys. Rev. Lett.}\ }\textbf {\bibinfo {volume} {110}},\ \bibinfo {pages}
  {234101} (\bibinfo {year} {2013})}\BibitemShut {NoStop}%
\bibitem [{\citenamefont {Cartarius}\ \emph {et~al.}(2012)\citenamefont
  {Cartarius}, \citenamefont {Haag}, \citenamefont {Dast},\ and\ \citenamefont
  {Wunner}}]{Cartarius12a}%
  \BibitemOpen
  \bibfield  {author} {\bibinfo {author} {\bibfnamefont {H.}~\bibnamefont
  {Cartarius}}, \bibinfo {author} {\bibfnamefont {D.}~\bibnamefont {Haag}},
  \bibinfo {author} {\bibfnamefont {D.}~\bibnamefont {Dast}}, \ and\ \bibinfo
  {author} {\bibfnamefont {G.}~\bibnamefont {Wunner}},\ }\href {\doibase
  10.1088/1751-8113/45/44/444008} {\bibfield  {journal} {\bibinfo  {journal}
  {J. Phys. A}\ }\textbf {\bibinfo {volume} {45}},\ \bibinfo {pages} {444008}
  (\bibinfo {year} {2012})}\BibitemShut {NoStop}%
\bibitem [{\citenamefont {Cartarius}\ and\ \citenamefont
  {Wunner}(2012)}]{Cartarius12b}%
  \BibitemOpen
  \bibfield  {author} {\bibinfo {author} {\bibfnamefont {H.}~\bibnamefont
  {Cartarius}}\ and\ \bibinfo {author} {\bibfnamefont {G.}~\bibnamefont
  {Wunner}},\ }\href {\doibase 10.1103/physreva.86.013612} {\bibfield
  {journal} {\bibinfo  {journal} {Phys. Rev. A}\ }\textbf {\bibinfo {volume}
  {86}},\ \bibinfo {pages} {013612} (\bibinfo {year} {2012})}\BibitemShut
  {NoStop}%
\bibitem [{\citenamefont {Dast}\ \emph
  {et~al.}(2013{\natexlab{a}})\citenamefont {Dast}, \citenamefont {Haag},
  \citenamefont {Cartarius}, \citenamefont {Wunner}, \citenamefont {Eichler},\
  and\ \citenamefont {Main}}]{Dast13a}%
  \BibitemOpen
  \bibfield  {author} {\bibinfo {author} {\bibfnamefont {D.}~\bibnamefont
  {Dast}}, \bibinfo {author} {\bibfnamefont {D.}~\bibnamefont {Haag}}, \bibinfo
  {author} {\bibfnamefont {H.}~\bibnamefont {Cartarius}}, \bibinfo {author}
  {\bibfnamefont {G.}~\bibnamefont {Wunner}}, \bibinfo {author} {\bibfnamefont
  {R.}~\bibnamefont {Eichler}}, \ and\ \bibinfo {author} {\bibfnamefont
  {J.}~\bibnamefont {Main}},\ }\href {\doibase 10.1002/prop.201200080}
  {\bibfield  {journal} {\bibinfo  {journal} {Fortschr. Physik}\ }\textbf
  {\bibinfo {volume} {61}},\ \bibinfo {pages} {124} (\bibinfo {year}
  {2013}{\natexlab{a}})}\BibitemShut {NoStop}%
\bibitem [{\citenamefont {Dast}\ \emph
  {et~al.}(2013{\natexlab{b}})\citenamefont {Dast}, \citenamefont {Haag},
  \citenamefont {Cartarius}, \citenamefont {Main},\ and\ \citenamefont
  {Wunner}}]{Dast13b}%
  \BibitemOpen
  \bibfield  {author} {\bibinfo {author} {\bibfnamefont {D.}~\bibnamefont
  {Dast}}, \bibinfo {author} {\bibfnamefont {D.}~\bibnamefont {Haag}}, \bibinfo
  {author} {\bibfnamefont {H.}~\bibnamefont {Cartarius}}, \bibinfo {author}
  {\bibfnamefont {J.}~\bibnamefont {Main}}, \ and\ \bibinfo {author}
  {\bibfnamefont {G.}~\bibnamefont {Wunner}},\ }\href {\doibase
  10.1088/1751-8113/46/37/375301} {\bibfield  {journal} {\bibinfo  {journal}
  {J. Phys. A}\ }\textbf {\bibinfo {volume} {46}},\ \bibinfo {pages} {375301}
  (\bibinfo {year} {2013}{\natexlab{b}})}\BibitemShut {NoStop}%
\bibitem [{\citenamefont {Haag}\ \emph {et~al.}(2014)\citenamefont {Haag},
  \citenamefont {Dast}, \citenamefont {L\"ohle}, \citenamefont {Cartarius},
  \citenamefont {Main},\ and\ \citenamefont {Wunner}}]{Haag14b}%
  \BibitemOpen
  \bibfield  {author} {\bibinfo {author} {\bibfnamefont {D.}~\bibnamefont
  {Haag}}, \bibinfo {author} {\bibfnamefont {D.}~\bibnamefont {Dast}}, \bibinfo
  {author} {\bibfnamefont {A.}~\bibnamefont {L\"ohle}}, \bibinfo {author}
  {\bibfnamefont {H.}~\bibnamefont {Cartarius}}, \bibinfo {author}
  {\bibfnamefont {J.}~\bibnamefont {Main}}, \ and\ \bibinfo {author}
  {\bibfnamefont {G.}~\bibnamefont {Wunner}},\ }\href {\doibase
  10.1103/PhysRevA.89.023601} {\bibfield  {journal} {\bibinfo  {journal} {Phys.
  Rev. A}\ }\textbf {\bibinfo {volume} {89}},\ \bibinfo {pages} {023601}
  (\bibinfo {year} {2014})}\BibitemShut {NoStop}%
\bibitem [{\citenamefont {{Achilleos, V. and Kevrekidis, P. G. and
  Frantzeskakis, D. J. and Carretero-Gonz\'alez, R.}}(2012)}]{Achilleos12a}%
  \BibitemOpen
  \bibfield  {author} {\bibinfo {author} {\bibnamefont {{Achilleos, V. and
  Kevrekidis, P. G. and Frantzeskakis, D. J. and Carretero-Gonz\'alez, R.}}},\
  }\href {\doibase 10.1103/PhysRevA.86.013808} {\bibfield  {journal} {\bibinfo
  {journal} {Phys. Rev. A}\ }\textbf {\bibinfo {volume} {86}},\ \bibinfo
  {pages} {013808} (\bibinfo {year} {2012})}\BibitemShut {NoStop}%
\bibitem [{\citenamefont {Konotop}\ \emph {et~al.}(2016)\citenamefont
  {Konotop}, \citenamefont {Yang},\ and\ \citenamefont
  {Zezyulin}}]{Konotop16a}%
  \BibitemOpen
  \bibfield  {author} {\bibinfo {author} {\bibfnamefont {V.~V.}\ \bibnamefont
  {Konotop}}, \bibinfo {author} {\bibfnamefont {J.}~\bibnamefont {Yang}}, \
  and\ \bibinfo {author} {\bibfnamefont {D.~A.}\ \bibnamefont {Zezyulin}},\
  }\href {\doibase 10.1103/RevModPhys.88.035002} {\bibfield  {journal}
  {\bibinfo  {journal} {Rev. Mod. Phys.}\ }\textbf {\bibinfo {volume} {88}},\
  \bibinfo {pages} {035002} (\bibinfo {year} {2016})}\BibitemShut {NoStop}%
\bibitem [{\citenamefont {{Schwarz, Lukas and Cartarius, Holger and Musslimani,
  Ziad H. and Main, J\"org and Wunner, G\"unter}}(2017)}]{Schwarz17a}%
  \BibitemOpen
  \bibfield  {author} {\bibinfo {author} {\bibnamefont {{Schwarz, Lukas and
  Cartarius, Holger and Musslimani, Ziad H. and Main, J\"org and Wunner,
  G\"unter}}},\ }\href {\doibase 10.1103/PhysRevA.95.053613} {\bibfield
  {journal} {\bibinfo  {journal} {Phys. Rev. A}\ }\textbf {\bibinfo {volume}
  {95}},\ \bibinfo {pages} {053613} (\bibinfo {year} {2017})}\BibitemShut
  {NoStop}%
\bibitem [{\citenamefont {Landau}\ and\ \citenamefont
  {Pitaevskii}(1979)}]{Landau79a}%
  \BibitemOpen
  \bibfield  {author} {\bibinfo {author} {\bibfnamefont {L.~D.}\ \bibnamefont
  {Landau}}\ and\ \bibinfo {author} {\bibfnamefont {L.~P.}\ \bibnamefont
  {Pitaevskii}},\ }\href@noop {} {\emph {\bibinfo {title} {Statistical Physics,
  Part 2: Theory of the Condensed State}}}\ (\bibinfo  {publisher} {Pergamon
  Press Ltd.},\ \bibinfo {address} {Oxford},\ \bibinfo {year}
  {1979})\BibitemShut {NoStop}%
\bibitem [{\citenamefont {Lin}\ \emph {et~al.}(2009)\citenamefont {Lin},
  \citenamefont {Compton}, \citenamefont {Jimenez-Garcia}, \citenamefont
  {Porto},\ and\ \citenamefont {Spielman}}]{Lin09a}%
  \BibitemOpen
  \bibfield  {author} {\bibinfo {author} {\bibfnamefont {Y.-J.}\ \bibnamefont
  {Lin}}, \bibinfo {author} {\bibfnamefont {R.~L.}\ \bibnamefont {Compton}},
  \bibinfo {author} {\bibfnamefont {K.}~\bibnamefont {Jimenez-Garcia}},
  \bibinfo {author} {\bibfnamefont {J.~V.}\ \bibnamefont {Porto}}, \ and\
  \bibinfo {author} {\bibfnamefont {I.~B.}\ \bibnamefont {Spielman}},\ }\href
  {\doibase 10.1038/nature08609} {\bibfield  {journal} {\bibinfo  {journal}
  {Nature}\ }\textbf {\bibinfo {volume} {462}},\ \bibinfo {pages} {628}
  (\bibinfo {year} {2009})}\BibitemShut {NoStop}%
\bibitem [{\citenamefont {Zhao}\ and\ \citenamefont {Gu}(2015)}]{Zhao15a}%
  \BibitemOpen
  \bibfield  {author} {\bibinfo {author} {\bibfnamefont {Q.}~\bibnamefont
  {Zhao}}\ and\ \bibinfo {author} {\bibfnamefont {Q.}~\bibnamefont {Gu}},\
  }\href {\doibase 10.1007/s11467-015-0505-x} {\bibfield  {journal} {\bibinfo
  {journal} {Frontiers of Physics}\ }\textbf {\bibinfo {volume} {10}},\
  \bibinfo {pages} {100306} (\bibinfo {year} {2015})}\BibitemShut {NoStop}%
\bibitem [{\citenamefont {Haag}\ \emph {et~al.}(2015)\citenamefont {Haag},
  \citenamefont {Dast}, \citenamefont {Cartarius},\ and\ \citenamefont
  {Wunner}}]{Haag15a}%
  \BibitemOpen
  \bibfield  {author} {\bibinfo {author} {\bibfnamefont {D.}~\bibnamefont
  {Haag}}, \bibinfo {author} {\bibfnamefont {D.}~\bibnamefont {Dast}}, \bibinfo
  {author} {\bibfnamefont {H.}~\bibnamefont {Cartarius}}, \ and\ \bibinfo
  {author} {\bibfnamefont {G.}~\bibnamefont {Wunner}},\ }\href {\doibase
  10.1007/s10773-014-2481-2} {\bibfield  {journal} {\bibinfo  {journal} {Int.
  J. Theor. Phys.}\ }\textbf {\bibinfo {volume} {54}},\ \bibinfo {pages} {4100}
  (\bibinfo {year} {2015})}\BibitemShut {NoStop}%
\bibitem [{\citenamefont {Anderson}\ \emph {et~al.}(2000)\citenamefont
  {Anderson}, \citenamefont {Haljan}, \citenamefont {Wieman},\ and\
  \citenamefont {Cornell}}]{Anderson00a}%
  \BibitemOpen
  \bibfield  {author} {\bibinfo {author} {\bibfnamefont {B.~P.}\ \bibnamefont
  {Anderson}}, \bibinfo {author} {\bibfnamefont {P.~C.}\ \bibnamefont
  {Haljan}}, \bibinfo {author} {\bibfnamefont {C.~E.}\ \bibnamefont {Wieman}},
  \ and\ \bibinfo {author} {\bibfnamefont {E.~A.}\ \bibnamefont {Cornell}},\
  }\href {\doibase 10.1103/PhysRevLett.85.2857} {\bibfield  {journal} {\bibinfo
   {journal} {Phys. Rev. Lett.}\ }\textbf {\bibinfo {volume} {85}},\ \bibinfo
  {pages} {2857} (\bibinfo {year} {2000})}\BibitemShut {NoStop}%
\bibitem [{\citenamefont {Kevrekidis}\ \emph {et~al.}(2003)\citenamefont
  {Kevrekidis}, \citenamefont {Carretero-Gonz\'alez}, \citenamefont
  {Theocharis}, \citenamefont {Frantzeskakis},\ and\ \citenamefont
  {Malomed}}]{Kevrekidis03a}%
  \BibitemOpen
  \bibfield  {author} {\bibinfo {author} {\bibfnamefont {P.~G.}\ \bibnamefont
  {Kevrekidis}}, \bibinfo {author} {\bibfnamefont {R.}~\bibnamefont
  {Carretero-Gonz\'alez}}, \bibinfo {author} {\bibfnamefont {G.}~\bibnamefont
  {Theocharis}}, \bibinfo {author} {\bibfnamefont {D.~J.}\ \bibnamefont
  {Frantzeskakis}}, \ and\ \bibinfo {author} {\bibfnamefont {B.~A.}\
  \bibnamefont {Malomed}},\ }\href@noop {} {\bibfield  {journal} {\bibinfo
  {journal} {J. Phys. B}\ }\textbf {\bibinfo {volume} {36}},\ \bibinfo {pages}
  {3467} (\bibinfo {year} {2003})}\BibitemShut {NoStop}%
\bibitem [{\citenamefont {Frantzeskakis}\ \emph {et~al.}(2002)\citenamefont
  {Frantzeskakis}, \citenamefont {Theocharis}, \citenamefont {Diakonos},
  \citenamefont {Schmelcher},\ and\ \citenamefont
  {Kivshar}}]{Frantzeskakis02a}%
  \BibitemOpen
  \bibfield  {author} {\bibinfo {author} {\bibfnamefont {D.~J.}\ \bibnamefont
  {Frantzeskakis}}, \bibinfo {author} {\bibfnamefont {G.}~\bibnamefont
  {Theocharis}}, \bibinfo {author} {\bibfnamefont {F.~K.}\ \bibnamefont
  {Diakonos}}, \bibinfo {author} {\bibfnamefont {P.}~\bibnamefont
  {Schmelcher}}, \ and\ \bibinfo {author} {\bibfnamefont {Y.~S.}\ \bibnamefont
  {Kivshar}},\ }\href {\doibase 10.1103/PhysRevA.66.053608} {\bibfield
  {journal} {\bibinfo  {journal} {Phys. Rev. A}\ }\textbf {\bibinfo {volume}
  {66}},\ \bibinfo {pages} {053608} (\bibinfo {year} {2002})}\BibitemShut
  {NoStop}%
\end{thebibliography}
\end{document}